\documentclass[conference]{IEEEtran}
\IEEEoverridecommandlockouts

\usepackage{cite}
\usepackage{amsmath,amssymb,amsfonts}
\usepackage{algorithmic}
\usepackage{graphicx}
\usepackage{textcomp}
\usepackage{xcolor}
\def\BibTeX{{\rm B\kern-.05em{\sc i\kern-.025em b}\kern-.08em
    T\kern-.1667em\lower.7ex\hbox{E}\kern-.125emX}}

\usepackage{soul}
\usepackage{listings}
\usepackage[symbol]{footmisc}
\sethlcolor{green}
\usepackage{subcaption}
\usepackage{makecell}
\usepackage{circledsteps}

\newcommand{\ignore}[1]{}

\newcommand\FilledCircle[2][]{
  \ifmmode
    \Circled[fill color=black,inner color=white,#1]{\mathsf{#2}}
  \else
    \Circled[fill color=black,inner color=white,#1]{\sffamily#2}
  \fi
}

\begin{document}

\title{Polaris: Multi-Fidelity Design Space Exploration of Deep Learning
Accelerators\\
\thanks{This work was funded by a gift from ARM and by an NXP fellowship.  We
thank Kevin Swersky for his collaboration on foundational concepts that this
paper builds on, and we thank Minesh Patel, Molly O'Neil, and Quang Duong for
their comments on early drafts of this paper.} }

\author{\IEEEauthorblockN{Chirag Sakhuja}
\IEEEauthorblockA{\textit{The University of Texas at Austin}\\
Austin, TX, USA \\
chirag.sakhuja@utexas.edu}
\and
\IEEEauthorblockN{Charles Hong}
\IEEEauthorblockA{\textit{University of California, Berkeley} \\
Berkeley, CA, USA \\
charleshong@berkeley.edu}
\and
\IEEEauthorblockN{Calvin Lin}
\IEEEauthorblockA{\textit{The University of Texas at Austin} \\
Austin, TX, USA \\
lin@cs.utexas.edu}
}

\maketitle

\begin{abstract}

This paper presents a tool for automatically exploring the design space of
deep learning accelerators (DLAs).

Our main advancement is Starlight, a data-driven performance model that uses
transfer learning to bridge the gap between fast, low-fidelity evaluation
methods (such as analytical models) and slow, high-fidelity evaluation
methods (such as RTL simulation).  Starlight is fast:  It can provide 6,500
predictions per second, allowing the evaluation of millions of configurations
per hour.  Starlight is accurate:  It predicts the energy-delay product
measured by RTL simulation with 99\% accuracy.  And Starlight can be trained
efficiently:  It can be trained with 61\% fewer samples than DOSA's
state-of-the-art data-driven performance predictor~\cite{hong2023dosa}.

Our second contribution is Polaris, a design-space exploration tool that
uses Starlight to efficiently search the large, complex hardware/software
co-design space of DLAs.  In under 35 minutes, Polaris produces DLA designs
that match the performance of designs that take six hours to produce with
DOSA.  And in under 3.3 hours, Polaris produces DLA designs that reduce
energy-delay product by 2.7$\times$ over the best designs found by DOSA.

\end{abstract}


\section{Introduction}
\label{sec:intro}

Because of the far-reaching impact of deep
learning~\cite{dong2021survey,maslej2023ai},
modern hardware systems often incorporate deep learning accelerators
(DLAs)~\cite{abts2022challenges,emani2021accelerating,jouppi2021ten,lavely2022powering,seshadri2022evaluation,sipola2022artificial,vasiljevic2021compute},
which for deep learning workloads are more energy- and area-efficient
than CPUs and GPUs~\cite{dhilleswararao2022efficient}.
Because of the fast pace of innovation in deep learning
models~\cite{chen2019eyeriss,jouppi2021ten,seshadri2022evaluation}, new
DLAs are constantly being developed.

One method of reducing the high
cost~\cite{jouppi2017indatacenter,lattner2021mlir}
of DLA design is to perform automated
design space exploration (DSE) of the hardware/software (HW/SW) co-design
space~\cite{dave2019dmazerunner,hong2023dosa,huang2022learning,kao2020confuciux,kumar2021datadriven,lin2021naas,mei2021zigzag,nardi2019practical,sakhuja2023leveraging,samajdar2023airchitect,venkatesan2019magnet,xiao2021hasco,yang2020interstellar,yazdanbakhsh2020apollo,zhang2022fullstack}.  This design space
consists of design parameters, such as spatial array dimensions, memory
sizes, and loop tiling factors.  DSE iteratively identifies points in the
design space and evaluates their quality based on some optimization criteria
(e.g., minimized delay or energy-delay product).

\begin{figure}[t]
    \centering
    \includegraphics[width=\linewidth]{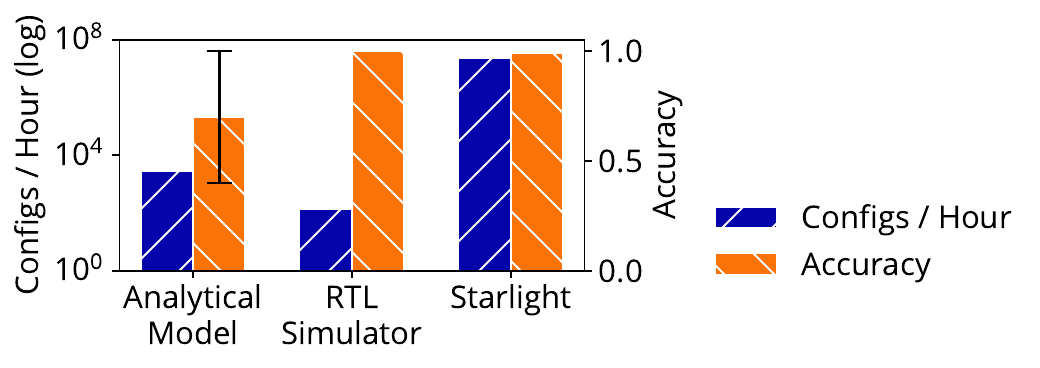}

    \caption{Analytical models can be queried thousands of times an hour, but
    they are inaccurate, whereas an an RTL simulator is accurate but slow.  Our
    learned model, Starlight, breaks this tradeoff by predicting performance
    faster than an analytical model and with 99\% accuracy when
    compared to an RTL simulator.  These data are collected by Parashar et
    al.~\cite{parashar2019timeloop}, Karandikar et
    al.~\cite{karandikar2018firesim}, and M\~{u}noz-Martinez et
    al.~\cite{munoz-martinez2021stonne}.}

    \label{fig:dilemma}

\end{figure}

For DSE tools, the choice of evaluation mechanism presents a fundamental
tradeoff.  The tool could evaluate configurations with high fidelity,
e.g., using an RTL simulator, but the long latency of such techniques would
restrict the DSE tool to considering just a small number of configurations,
severely limiting the tool's effectiveness.  Alternatively, the tool
could evaluate a large number of configurations using a fast evaluation
method, e.g., an analytical model, but such techniques have low fidelity
because they do not capture the nuances of circuitry or runtime behavior.
Figure~\ref{fig:dilemma} illustrates this tradeoff:  As model fidelity
increases, the number of configurations that a DSE tool can evaluate
dramatically decreases.

Prior work~\cite{hong2023dosa,kumar2021datadriven} attempts to break this
tradeoff by using a fast {\em data-driven model} that has been trained
to predict a design's performance as measured by a high-fidelity method,
such as RTL simulation. Such a data-driven model can be queried even
faster than an analytical model and can produce results that approach
the accuracy of RTL simulation.  Unfortunately, the training required to
produce such a data-driven model itself requires thousands of high-fidelity
evaluations~\cite{hong2023dosa,kumar2021datadriven}---which even as a
one-time investment is difficult to collect---presenting the same tradeoff
that the data-driven model was intended to break.

In this paper, we break this tradeoff by employing a technique called
{\it transfer learning} to more efficiently train a data-driven model.
Transfer learning uses as a starting point an existing data-driven
model that is trained on a similar but different prediction task.
We use transfer learning to train a data-driven model that predicts the
performance of a DLA as measured by RTL simulation; our model is trained
primarily on easy-to-collect, low-fidelity evaluations (analytical model)
rather than hard-to-collect, high-fidelity evaluations (RTL simulation).
We are the first to apply transfer learning  this way.

Figure~\ref{fig:dilemma} shows that our data-driven model, called Starlight,
can be queried faster than an analytical model and can achieve 99\% accuracy
when predicting the energy-delay product (EDP) of a DLA.  Moreover, Starlight
is trained with 61\% fewer high-fidelity evaluations and achieves higher
accuracy than DOSA's state-of-the-art data-driven model~\cite{hong2023dosa}.

\ignore{
---to reduce the number of slow evaluations necessary to train the latter.

in which a data-driven model that is trained to perform one
prediction task is used to reduce the training data necessary to perform a
similar but different prediction task.  In particular, we are the first to use
transfer learning to transfer knowledge from a low-fidelity data-driven
model---i.e., a model that is trained on evaluations by an analytical model---to
a high-fidelity data-driven model---i.e., a model that is trained on evaluations
by RTL simulation---to reduce the number of slow evaluations necessary to train
the latter.
}

Because Starlight achieves such high accuracy, we might be tempted to
perform DSE with Starlight using the {\it offline} approach that prior work
has taken~\cite{kumar2021datadriven}, namely, perform optimization (e.g.,
stocastic gradient descent) on Starlight and evaluate the final resulting
design with RTL simulation.  But even a highly accurate model may not capture
details of the actual hardware, so a design that is deemed high-quality by the
model might not be high-quality when translated to real hardware.  Thus, it
might be necessary to perform RTL simulation {\it in} the optimization loop,
which is known as {\it online} DSE.  Others have suggested that offline
approaches are sufficient~\cite{kumar2021datadriven}, but in this paper, we show
for the first time that there {\em is} significant advantage to building an
online DSE tool to ensure that the designs are faithful when translated to real
hardware.

\ignore{
Because Starlight achieves such high accuracy, we might be tempted to
perform DSE with Starlight using the {\it offline} approach that prior
work has taken~\cite{kumar2021datadriven}, namely, perform optimization
on Starlight and only evaluate the final resulting design with RTL
simulation.  In fact, others have suggested that such offline approaches
are sufficient~\cite{kumar2021datadriven}.  But even a highly accurate
model ignores details of the actual hardware, so a design that is deemed
high-quality by the model might not be high-quality when translated to
real hardware.  In this paper, we show for the first time that
there {\em is} significant advantage to building a DSE tool that performs
RTL simulation inside the optimization loop---i.e., it performs {\it online}
DSE to ensure that the designs are faithful when translated to real hardware.
}

We do so by building Polaris, a DSE tool that integrates Starlight into a
Bayesian optimization (BO) framework.  BO is sample-efficient because it
carefully selects the designs that should be evaluated using RTL simulation.
In particular, Polaris uses Starlight to balance the exploitation of promising
regions of the co-design space with the exploration of uncertain regions
of the co-design space.  Because Polaris utilizes a mixture of low- and
high-fidelity evaluations, we say that it performs multi-fidelity design space
exploration.

This paper makes the following contributions:

\begin{itemize}

    \item We demonstrate that transfer learning is an effective method of
	  building data-driven performance models.  Our model is trained
	  with 61\% fewer evaluations than DOSA's state-of-the-art
	  data-driven model~\cite{hong2023dosa}.

    \item We present Starlight, a data-driven model that predicts with
          99\% accuracy the EDP of a DLA as measured by RTL simulation.
          Transfer learning allows Starlight to be trained in 2 minutes
          on a consumer-grade CPU.

    \item We demonstrate the benefit of performing RTL simulation in the
          optimization loop by presenting Polaris, a DSE
          tool that integrates Starlight into a Bayesian optimization
          framework to perform hardware/software co-design of DLAs.
          Polaris produces in 35 minutes DLA designs and software
          mappings that have lower EDP than those produced in 6 hours
          by DOSA~\cite{hong2023dosa}, which uses an offline approach.
          And within 3.3 hours, Polaris' designs achieve an average
          reduction of 2.7$\times$ in EDP over the best designs produced
          by DOSA.

\end{itemize}

The remainder of this paper is organized as follows. Section~\ref{sec:related}
places this work in the context of prior work. Section~\ref{sec:background} then
provides background for understanding our solution.  Section~\ref{sec:starlight}
and Section~\ref{sec:polaris} describe in greater detail Starlight and Polaris,
respectively. Sections~\ref{sec:methodology} and \ref{sec:evaluation} present
the evaluation methodology and evaluation results of Starlight and Polaris, and
Section~\ref{sec:conclusion} provides concluding remarks.

\section{Related Work}
\label{sec:related}

This section places our work in the context of prior work.

\subsection{Performance Predictors}
Data-driven models are a valuable tool because they can predict the performance
of a DLA with accuracy that approaches RTL simulation, but they can be queried
orders of magnitude more quickly than RTL simulation.

Kaufman et al.~\cite{kaufman2021learned} design the first general-purpose
data-driven model, which is a graph neural network that estimates delay by
consuming as input a tensor computation graph and DLA-specific opcodes.
Esmaeilzadeh et al.~\cite{esmaeilzadeh2022physically} and Ferianc et
al.~\cite{ferianc2021improving} present data-driven models that take as input
abstracted convolutional layer shapes and/or DLA architectural parameters, but
they only support a subset of the design space that Starlight supports.
AIrchitect~\cite{samajdar2023airchitect} is a recommendation model that
automatically predicts for a given workload optimized design parameters, but
AIrchitect does not account for the DLA's architectural parameters. Other
data-driven models are integrated into DSE
tools~\cite{esmaeilzadeh2023opensource,hong2023dosa,huang2022learning,kumar2021datadriven}.

Some data-driven
models~\cite{esmaeilzadeh2022physically,hong2023dosa,kumar2021datadriven} are
trained with samples from a slow high-fidelity method, such as RTL simulation,
which are difficult to collect.  Starlight is the first data-driven model to
reduce the cost of training by transferring knowledge from a low-fidelity
data-driven performance model to a high-fidelity data-driven performance model.

\subsection{Automated HW/SW Co-Design of DLAs}
Recent work in the automated DSE of deep learning accelerators
(DLAs)~\cite{kim2023full,pham-quoc2021hardware,sekanina2021neural,talbi2021automated}
commonly targets the exploration of the joint design space of architectural
parameters---e.g.  systolic array size---and software mappings---e.g. loop
tiling factors---because the two design spaces are tightly coupled and are
profitable to explore simultaneously~\cite{parashar2019timeloop,shao2019simba}.
This type of DSE is called HW/SW co-design.

For modest co-design spaces, simple black-box optimization
algorithms~\cite{dave2019dmazerunner,mei2021zigzag} that require no knowledge of the
shape of the objective function are feasible, but larger
co-design spaces require more sophisticated techniques, such as Bayesian
optimization~\cite{nardi2019practical,sakhuja2023leveraging,venkatesan2019magnet,xiao2021hasco,yazdanbakhsh2020apollo,zhang2022fullstack},
reinforcement learning~\cite{kao2020confuciux,xiao2021hasco},
or genetic algorithms~\cite{lin2021naas,kao2020confuciux}.
Recent work uses algorithms, such as stochastic gradient descent, that optimize
a differentiable data-driven proxy
model~\cite{esmaeilzadeh2023opensource,hong2023dosa,huang2022learning,kumar2021datadriven}.

None of these tools perform RTL simulation in the
optimization loop.  Instead, they evaluate designs with
a performance model, and the final designs are evaluated with RTL
simulation~\cite{esmaeilzadeh2023opensource,hong2023dosa,kumar2021datadriven,koeplinger2018spatial,yazdanbakhsh2020apollo}.
But inaccuracies in the performance model lead to suboptimal designs
after they are translated to hardware.  Polaris avoids this inaccuracy by
performing RTL simulation in the optimization loop.

\section{Background}
\label{sec:background}

This section provides background material that is useful for understanding
our solution.  We first provide an overview of deep learning workloads
and deep learning accelerators.  We then explain Bayesian optimization
(BO), which is central to Polaris, before describing three relevant machine
learning techniques, namely, variational autoencoders, deep kernel learning,
and transfer learning.

\subsection{Deep Learning Workloads and Accelerators}

\begin{figure}[t]
    \centering
    \includegraphics[width=0.8\linewidth]{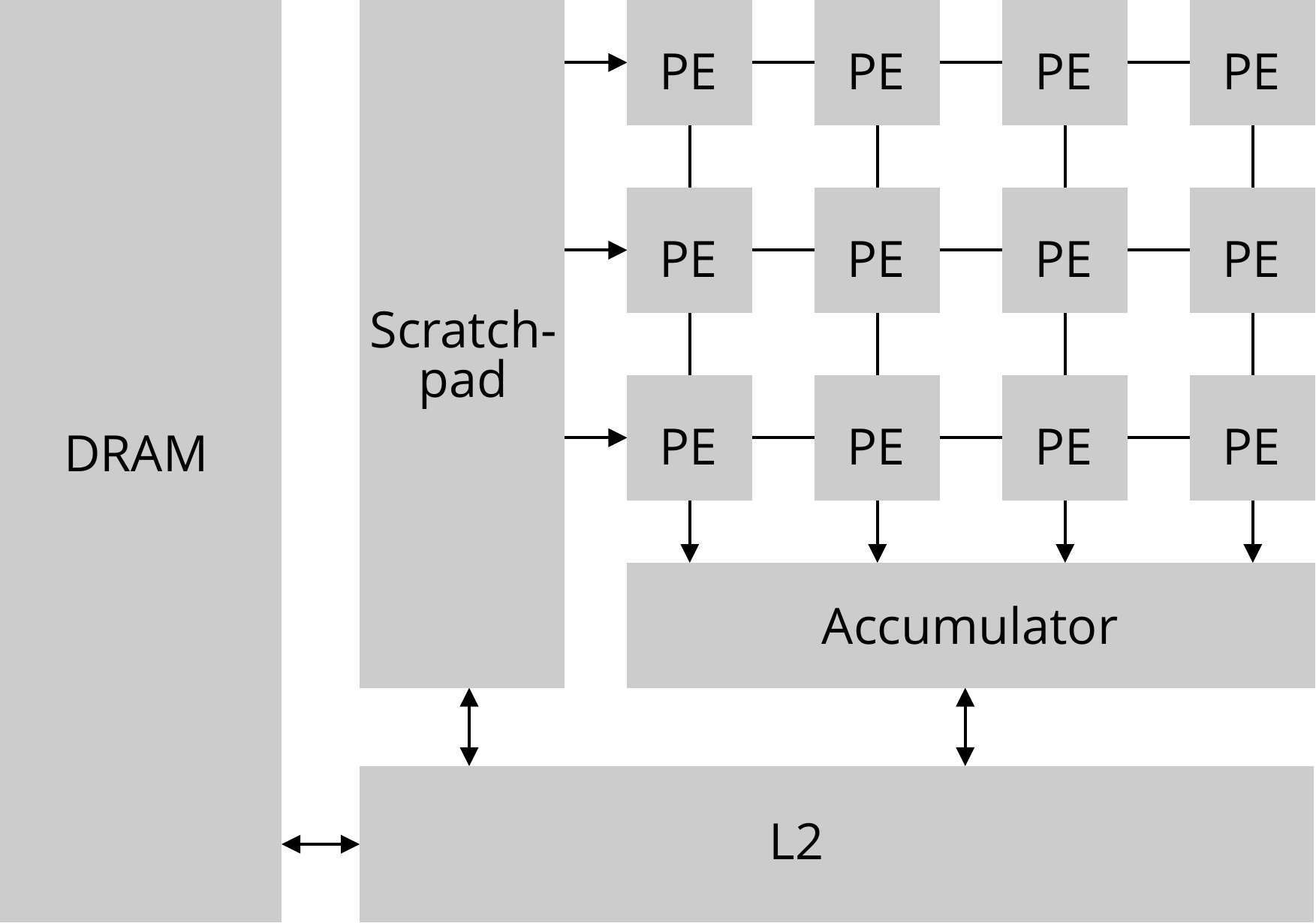}

    \caption{A typical deep learning accelerator.  Data is tiled in the
    scratchpad and fed into processing elements (PEs), which compute a
    convolution operation, using the accumulator to aggregate partial results.}

    \label{fig:accelerator}
\end{figure}

Deep learning (DL) models consist of a series of layers that can each be
represented by a convolution operation,\footnote{A convolution operation can
represent a matrix multiply (GEMM) operation without loss of generality.}
so they are common targets for acceleration.  A convolution is an operation
on $K$ weight tensors of size $R \times S \times C$ and $N$ input tensors
of size $(P + R - 1) \times (Q + S - 1) \times C$ to produce $N$ output
tensors of size $P \times Q \times K$.  The convolution is implemented
as a seven-level nested loop.  Additional parameters of the convolution
operation are the stride and dilation.

Deep learning accelerators (DLAs) execute the convolution operation
on a spatial array of processing elements (PEs) connected to a
specialized memory hierarchy that exploits the unique data reuse found
in a convolution~\cite{chen2017eyeriss}.  Figure~\ref{fig:accelerator}
shows the primary components of the DLA we use for this work,
Gemmini~\cite{genc2021gemmini}.  The PEs compute a partial convolution
operation and stores intermediate results in an on-chip memory structure called
an accumulator.  To compute an entire DL layer, Gemmini streams data from
DRAM, an L2 cache, and a software-managed scratchpad.

The data movement is precisely dictated by a per-layer {\em software mapping}.
Specifically, a software mapping describes (1) how each tensor is tiled to fit
in the memory hierarchy and (2) the order of the nested loops.

\subsection{Bayesian Optimization}
\label{sec:bayesian-optimization}

\begin{figure}[t]
    \centering
    \includegraphics[width=\linewidth]{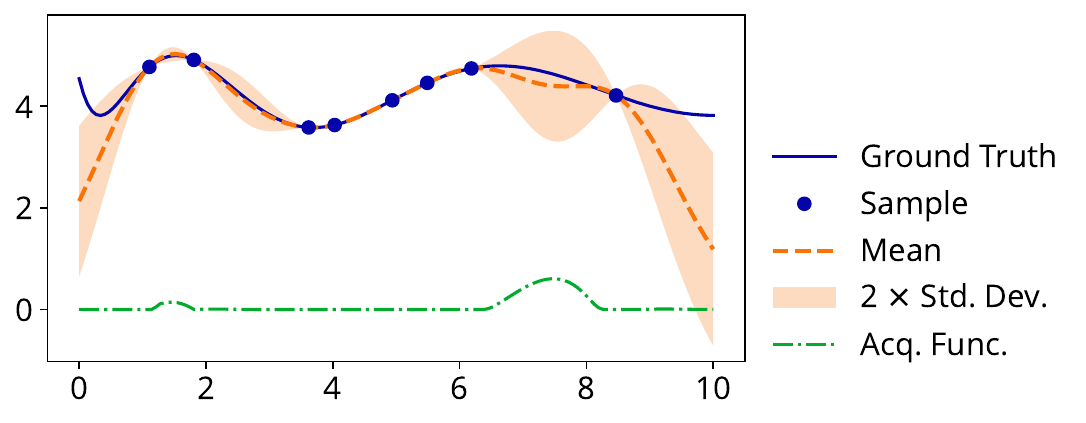}

    \caption{A Gaussian process that models a ground truth function that
    has been sampled at 8 points.  The acquisition function---in this
    case, Expected Improvement---is applied over the Gaussian process and
    maximized to determine the next sample to evaluate.}

    \label{fig:gp}
\end{figure}

\begin{figure*}[t]
    \centering
    \includegraphics[width=\linewidth]{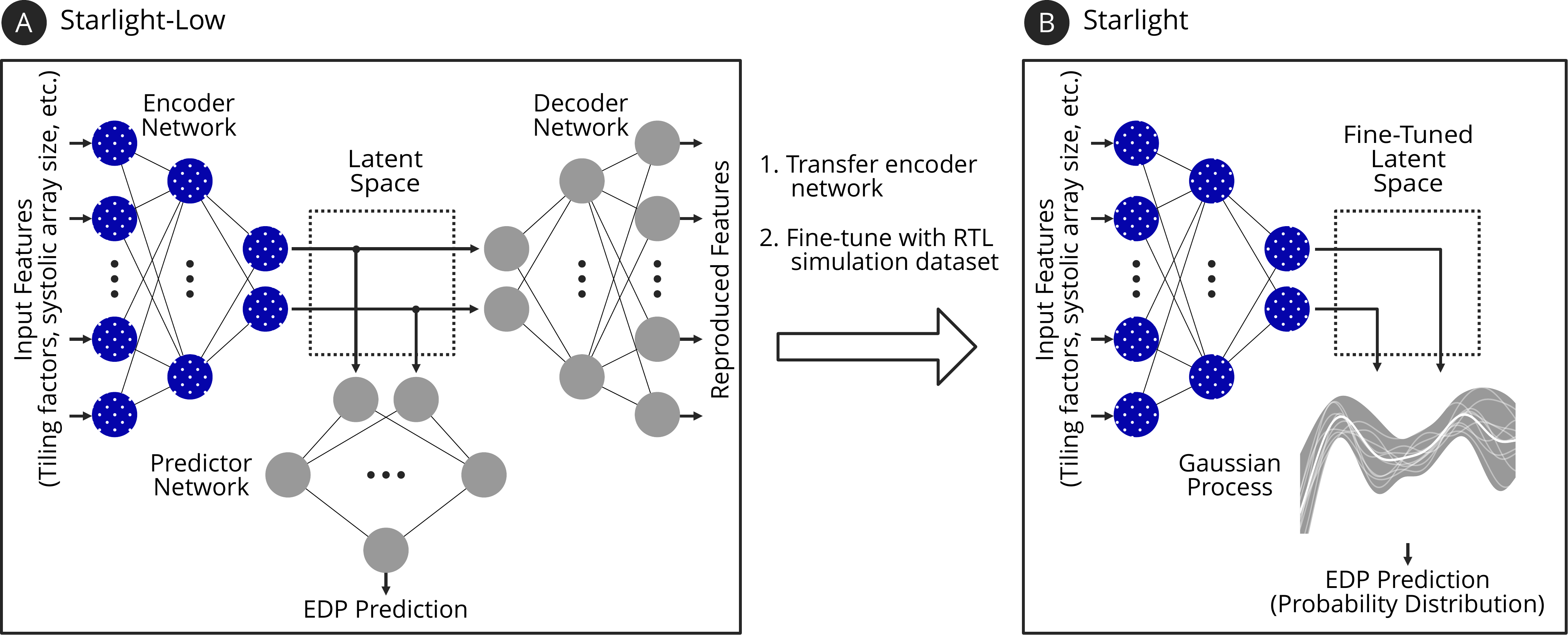}

    \caption{\FilledCircle{A} Starlight-Low is a neural network that predicts
    the energy-delay product (EDP) of a DLA as measured by a low-fidelity
    method, namely, an analytical model.  The encoder network (in blue dotted
    pattern) from Starlight-Low is transferred to \FilledCircle{B} Starlight,
    which is a machine learning model based on deep kernel learning that
    predicts the EDP as measured by a high-fidelity method, namely, an RTL
    simulator. The decoder network is dropped because it is no longer needed.}

    \label{fig:starlight}

\end{figure*}

{\em Bayesian optimization} (BO) is an optimization strategy that is commonly
employed when the {\em objective function}---i.e., the function that is being
optimized---is expensive to evaluate~\cite{bischl2023hyperparameter}.  BO has
been successfully applied to perform DSE of the hardware/software (HW/SW)
co-design space of
DLAs~\cite{nardi2019practical,reagen2017case,sakhuja2023leveraging,venkatesan2019magnet,xiao2021hasco,zhang2022fullstack}.
A BO framework consists of (1) a {\em surrogate model}, which cheaply predicts
the value of the objective function, and (2) an {\em acquisition function},
which selects the next sample that should be evaluated.

The surrogate model is a data-driven model that must maintain a reliable
measurement of uncertainty for its predictions---i.e., the output
is a probability distribution rather than a single prediction.
The most common type of surrogate model for BO is a Gaussian process
(GP)~\cite{forrester2007multifidelity}, which produces a Gaussian
distribution for a given input.  Figure~\ref{fig:gp} shows a GP (shaded
region and dashed line in orange) that is modeling a function (solid line
in blue) given 8 samples.  The shaded region represents the uncertainty
of the surrogate model at any given input.

The acquisition function is a function that is applied over the surrogate
model to balance exploration of uncertain regions with exploitation of
the regions that are likely to contain the optimum.  A common acquisition
function is Expected Improvement (EI), which measures, as a function of
every possible input, the change in expected value of the surrogate model.
The acquisition function is maximized to select the next sample that
should be evaluated.  Figure~\ref{fig:gp} shows the EI acquisition function
(dot-dash line in green) applied to the example GP.  The acquisition function
is maximized around $x = 7.5$, which is a region of high uncertainty.
After evaluating that point, the acquisition function will be maximized
around $x=1.5$, which is the maximum of the ground truth.

Given the surrogate model and acquisition function, BO repeatedly performs
the following operations, either for a fixed number of iterations or
until a stopping criterion is met:  (1) Select a sample by maximizing the
acquisition function, (2) evaluate the sample on the expensive objective
function, and (3) train the surrogate model with the new evaluation.

\subsection{Machine Learning Techniques to Predict Performance}
\label{sec:deep-kernel-learning}

The traditional method of evaluating
hardware is through the combined use of analytical
models~\cite{kwon2020maestro,parashar2019timeloop,samajdar2020systematic},
timing simulators~\cite{munoz-martinez2021stonne,xi2020smaug}, and
RTL simulators~\cite{karandikar2018firesim}, but recently a variety of
machine learning techniques have been used to predict the performance of
DLAs\cite{esmaeilzadeh2023opensource,hong2023dosa,krishnan2023archgym,kumar2021datadriven}.
Two techniques that have shown promising
results~\cite{bai2021boomexplorer,huang2022learning} are
autoencoders~\cite{kingma2022autoencoding,rezende2014stochastic} and deep
kernel learning~\cite{wilson2016deep}.

An autoencoder~\cite{rumelhart1986learning} is a type of DL model that learns
to compress with minimal loss a high-dimensional input into a low-dimensional
space called a {\it latent space}.  The architecture of an autoencoder is
shown in the top part of Figure~\ref{fig:starlight}~\FilledCircle{A}.  On the
left side of the autoencoder, in what is called the encoder network, is a
series of fully-connected layers that decrease in size until they reach the
target dimension of the latent space.  On the right side of the autoencoder,
in what is called the decoder network, is a series of fully-connected layers
that mirrors the architecture of the encoder network.  The autoencoder is
trained to minimize the loss between the input of the encoder network and
the output of the decoder network, which should precisely reconstruct the
original input.  Consequently, the autoencoder learns to encode inputs
into unique representations in the low-dimensional latent space.

To avoid overfitting~\cite{bank2023autoencoders}, we can inject randomness
into the latent representations.  Specifically, the last layer of the
encoder network is modified to output a Gaussian distribution---as
opposed to a scalar value---that non-deterministically encodes an
input into the latent space.  These so-called variational autoencoders
(VAEs)~\cite{kingma2022autoencoding} are widely accepted to be more robust
than standard autoencoders.

Autoencoders and VAEs can accurately predict the performance
of a DLA, and they can be trained with transfer learning
(Section~\ref{sec:transfer-learning}), but they are not suitable
for use within a BO framework alone because they do not provide a reliable
measurement of uncertainty.  By contrast, Gaussian processes (GPs)
do provide a reliable measurement of uncertainty, but GPs struggle to
make predictions from high-dimensional input spaces such as the HW/SW
co-design space~\cite{sakhuja2023leveraging}.  To build a model that
is suitable for BO---which is our intended use case for Starlight---we
employ a recent technique from prior work called deep kernel learning
(DKL)~\cite{bai2023transfer,feurer2022practical,li2023study,wistuba2020fewshot}.
DKL attaches an encoder network from a VAE to a Gaussian process (GP) to
overcome limitations of the individual techniques:  The encoder network
reduces the dimensionality of the input space, and the GP provides a
reliable measurement of uncertainty.

\subsection{Transfer Learning}
\label{sec:transfer-learning}

Transfer learning is a machine learning training technique that re-uses a
model for a different task than it was originally trained for.  There are
many forms of transfer learning~\cite{zhuang2021comprehensive}, but we
focus on a straightforward form called hard weight sharing that directly
transfers some trained weights from a {\em source model} to an untrained
{\em target model}.

\section{Starlight}
\label{sec:starlight}

\begin{figure}[t]
    \centering
    \begin{subfigure}[b]{0.45\linewidth}
        \centering
        \includegraphics[width=\textwidth]{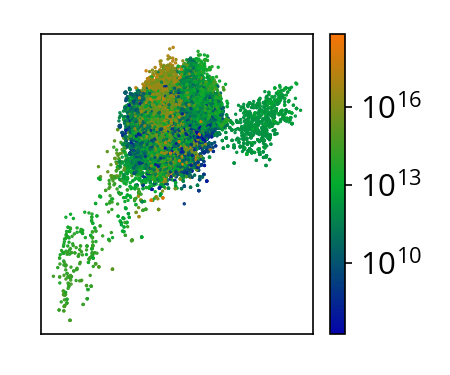}

        \caption{Without predictor network.}

        \label{fig:latent-space-without}
    \end{subfigure}
    \hspace{0.03\textwidth}
    \begin{subfigure}[b]{0.45\linewidth}
        \centering
        \includegraphics[width=\textwidth]{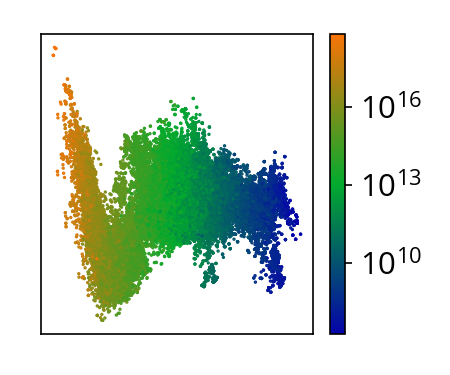}

        \caption{With predictor network.}

        \label{fig:latent-space-with}
    \end{subfigure}

    \caption{The 2-D latent space of a VAE trained (a) without a predictor
    network and (b) simultaneously with a predictor network.  Each point
    represents a HW/SW configuration that is color-coded by the EDP
    as measured by an analytical model. The predictor network induces
    structure, as indicated by the gradient of EDPs in the latent space.}

\end{figure}

In this section, we first present Starlight's inputs and outputs and the dataset
used for training.  We then show that transfer learning can be applied to
HW/SW co-design, and we present the source model, Starlight-Low, used to
transfer knowledge to Starlight. Finally, we present Starlight, which is an
accurate performance estimator that predicts the energy-delay product (EDP) of a
DLA as measured by RTL simulation.

\subsection{Inputs and Outputs}

The inputs to Starlight-Low and Starlight are (1) the architectural parameters
of a DLA and (2) the software mapping of a single convolutional layer.  The
output of Starlight is a Gaussian distribution that predicts energy-delay
product (EDP) such that the mean represents the prediction and the standard
deviation represents the uncertainty.

\begin{table}[t]
    \centering
    \begin{tabular}{|c|c|c|}
      \hline
      {\bf Parameter} & {\bf Values} \\
      \hline
      \hline
      Spatial Array Dimensions & 4x4, 8x8, 16x16, 32x32 \\
      \hline
      Accumulator Size & 8 to 256 KB (Step Size: 8) \\
      \hline
      Scratchpad Size & 8 to 256 KB (Step Size: 8) \\
      \hline
      Loop Order & Permutations of outermost loops \\
      \hline
      Tiling Factors\textsuperscript{\textdagger} & Divisors of layer shape \\
      \hline
    \end{tabular}

    \text{\textsuperscript{\textdagger}Independent values per level of memory
    hierarchy.}

    \caption{The ranges of parameter values in the input space of Starlight.}

    \label{tab:design-space}
\end{table}

The precise hardware and software design space that Starlight-Low and
Starlight are trained on is shown in Table~\ref{tab:design-space}.  In the
hardware design space, both models accept as input the spatial array size
and the accumulator and scratchpad sizes.  In the software design space,
both models accept as input the loop order and tiling factors.  The loop
order is encoded as a numerical value from 0 to 6 for each of the seven
loops in the convolutional layer loop nest.  All inputs are scaled to
the range $[0, 1]$ using a min-max scaler.

\subsection{Dataset}
\label{sec:starlight-dataset}

To train Starlight-Low, we collect a dataset of samples from the popular
Timeloop analytical model~\cite{parashar2019timeloop}, and to train
Starlight, and we collect a dataset of samples from the FireSim RTL
simulator~\cite{karandikar2018firesim}.  The datasets are collected by
performing Sobol sampling~\cite{sobol1967distribution} cut for
space: ---a sampling method
that results in a balanced dataset~\cite{esmaeilzadeh2023opensource}---on
the input space.  We collect a total of $2^{16}$ samples from Timeloop
and $2^{12}$ samples from FireSim.

We use both Timeloop and FireSim to measure the performance of the Gemmini
DLA~\cite{genc2021gemmini} when executing individual layers from one of
four DL models (see Section~\ref{sec:evaluation}).

A limitation of our training data, and consequently of Starlight,
is that FireSim does not measure energy consumption, so like prior
work~\cite{hong2023dosa}, we measure energy consumption using Timeloop.
For the remainder of this paper, EDP refers to the product of
energy consumption as measured by Timeloop and delay as measured by
FireSim.\footnote{To ensure the shared energy consumption measurement from
Timeloop is not contaminating Starlight,  we reproduce all experiments
using only delay measurements, which are independently measured by Timeloop
and FireSim.  The behavior is identical to the experiments that use EDP.}

\subsection{Applicability of Transfer Learning}

\begin{figure}[t]
    \centering
    \includegraphics[width=\linewidth]{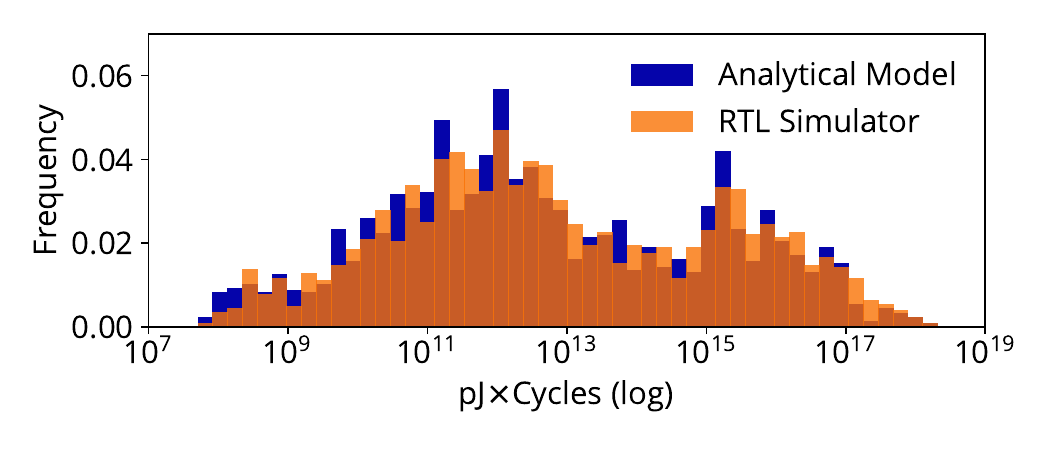}

    \caption{The distribution of EDPs for the same $2^{12}$ HW/SW configurations
    as measured by an analytical model and by an RTL simulator. The similarity
    of the distributions indicates that knowledge can be transferred between
    models.}

    \label{fig:transfer-learning-hist}
\end{figure}

Transfer learning can be applied when the information used to predict one task
can be transferred to the prediction of a different task.  To measure the
transferability of information between the Timeloop and FireSim datasets.
Figure~\ref{fig:transfer-learning-hist} shows the two distributions of EDPs.
We compute the Kullback-Leibler (KL) divergence~\cite{kullback1951information}---a
common metric to measure distribution similarity (lower means more similar).
For the overall distribution, the KL divergence is 0.04, indicating that
information can be transferred.  Furthermore, the KL divergence of the designs
with the lowest 10\% of EDPs---i.e., the key designs for design space
exploration---is 0.12, indicating that there is information in the target
distribution that a transferred model must learn to predict.

\subsection{Starlight-Low}

Starlight-Low is a neural network that predicts the EDP
of Gemmini as measured by a low-fidelity method, namely,
Timeloop~\cite{parashar2019timeloop}.  Starlight-Low is used as the source
model to transfer weights to Starlight.

The model architecture for Starlight-Low is based on a variational autoencoder
(VAE) because VAEs reduce the dimensionality of the inputs, which is important
when we incorporate a Gaussian process in Section~\ref{sec:starlight-starlight}.
Traditionally, a VAE connects an encoder network to a symmetric decoder network
and is trained to make the output reproduce the input exactly.  We build and
train a traditional VAE that encodes inputs into a 2-D latent space, which is
shown in Figure~\ref{fig:latent-space-without}.  Each point represents a HW/SW
configuration, and the color indicates the EDP as measured by an analytical
model.  There is no apparent structure to the latent space, which indicates that
the encoder is not properly learning the semantics of the inputs.  Consequently,
the latent space cannot reliably be used to make EDP predictions.

To induce structure in the latent space, as shown by the smooth gradient of EDPs
in Figure~\ref{fig:latent-space-with}, prior
work~\cite{gomez-bombarelli2018automatic,huang2022learning} simultaneously
trains a predictor network alongside the encoder and decoder networks.
Figure~\ref{fig:starlight} \FilledCircle{A} shows the model architecture of
Starlight-Low, which implements this technique.  The inputs are encoded into the
latent space and then fed to two outputs: the predictor network, which predicts
the EDP of the configuration, and the decoder network, which reproduces the
inputs to ensure that significant information is not lost in the latent space.

The final architecture of Starlight-Low is precisely as follows.  The encoder
network comprises fully connected layers of sizes 40, 24, 12, and 2, and
the decoder network mirrors the encoder network.  The predictor network
comprises fully connected layers of sizes 2, 64, 256, 256, 64, and 1.  In all
cases, layers are fed through a ReLU activation function.

Starlight-Low is trained to minimize (1) the mean squared error between the
predicted EDP and ground truth EDP, (2) the mean squared error between the
reproduced inputs and actual inputs, and (3) the KL divergence between the
latent encoding and unit multivariate Gaussian distribution.  Minimizing KL
divergence is the standard approach to ensure that the VAE does not collapse to
a traditional autoencoder during training.

\subsection{Starlight}
\label{sec:starlight-starlight}
Starlight is a machine learning model that predicts the EDP of a DLA as measured
by a high-fidelity method, namely, an RTL simulator.  Because Starlight is
designed for use within a Bayesian optimization (BO) framework, it must provide
a reliable measurement of uncertainty.  To achieve this, we build Starlight
using a technique called deep kernel learning~\cite{wilson2016deep} that fuses a
neural network (which does not provide the measurement of uncertainty) with a
Gaussian process (does provide the measurement of uncertainty).

To transfer knowledge from Starlight-Low to Starlight, we directly transfer
the weights from the encoder network of Starlight-Low.  We then fine-tune
Starlight using a dataset of EDPs as measured by RTL simulation.
We empirically validate this application of transfer learning in
Section~\ref{sec:evaluation-robustness}.

To build Starlight, we modify the architecture of Starlight-Low in two
key ways.

First, we remove the decoder network, because its goal is to prevent
information loss in the latent space, but the encoder network that is
transferred from Starlight-Low to Starlight already produces a well-behaved
latent space.

Second, we replace the predictor network in Starlight-Low with a Gaussian
process (GP).  This neural model architecture, which ties together a
neural network and a GP, is known as deep kernel learning (DKL) and is
essential for enabling Starlight to be used as a surrogate model for a
BO framework.  DKL lends two additional benefits: (1) unlike a standalone
GP, which is the typical surrogate model for a BO framework, DKL supports
transfer learning, and (2) DKL trains more robustly than other approaches,
as shown in Section~\ref{sec:evaluation-robustness}.

The final architecture of Starlight is shown in Figure~\ref{fig:starlight}
\FilledCircle{B}.  The GP in Starlight uses a Mat\'{e}rn
kernel~\cite{matern1986spatial} and gamma prior.  To train Starlight,
we maximize the marginal log likelihood of the combined encoder and
GP~\cite{wilson2016deep}.

\section{Polaris}
\label{sec:polaris}

\begin{figure*}[t]
    \centering
    \includegraphics[width=0.9\linewidth]{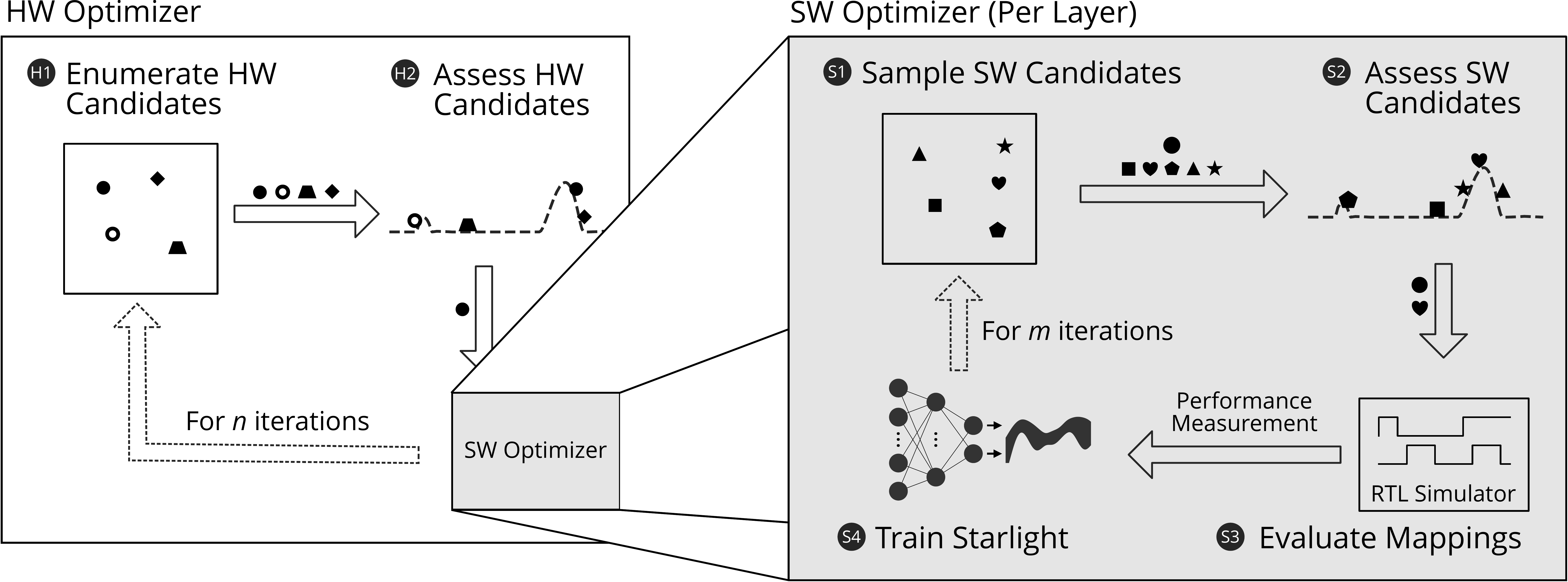}

    \caption{Polaris is a DSE tool that takes as input layer shapes
    that define the workload to be optimized and outputs an optimized
    DLA and software mappings.  The optimizer is split into an outer
    loop to optimize hardware and an inner loop to optimize software.
    The sequence of operations is as follows.  A hardware candidate
    is selected (\FilledCircle{H1}) and rounded to the nearest---as
    measured by distance in the latent space---implementable configurations
    (\FilledCircle{H2}).  Then, software candidates are selected for every
    layer (\FilledCircle{S1}) and rounded to the nearest implementable
    configurations (\FilledCircle{S2}) before being evaluated with an RTL
    simulator (\FilledCircle{S3}).  Finally, Starlight is updated with
    the new evaluations (\FilledCircle{S4}).  The process repeats for
    {\it n} trials in the hardware optimizer and {\it m} trials in the
    software optimizer.}

    \label{fig:polaris}

\end{figure*}

Polaris is a Bayesian optimization (BO) framework built around Starlight that
explores the co-design space of DLA design parameters and software mappings.
Specifically, the inputs to Polaris are the shapes of the convolutional
layers to be optimized; the outputs are the (1) architectural parameters
for a DLA and (2) software mappings that minimize the energy-delay product
(EDP) measured by RTL simulation.  Polaris uses Starlight as its surrogate
model, and it uses Upper Confidence Bound~\cite{srinivas2010gaussian}
as its acquisition function.

In this section, we first describe Polaris' iterative HW-SW
design process.  We then describe the HW and SW optimizers.

\subsection{Iterative Hardware-Software Design}

It is challenging for a HW/SW co-design tool to simultaneously co-design both
the hardware design and software mappings for all layers of a model because
the co-design space is enormous:  It is the Cartesian product of all hardware
and software design parameters, e.g., $O(10^{140})$ for ResNet-50, which
is a neural network used for image classification.  Thus, similar to prior
work~\cite{huang2022learning,lin2021naas,sakhuja2023leveraging,venkatesan2019magnet,xiao2021hasco,zhang2022fullstack},
Polaris is built using an iterative approach;  it first selects a hardware
candidate, then it optimizes each layer individually to find a software
mapping that minimizes the EDP of running that layer on the selected hardware
candidate.  Figure~\ref{fig:polaris} shows an overview of our approach.

\subsection{Hardware Optimizer}

The first step in an iteration of optimization with Polaris is to
select a hardware candidate.  In a traditional Bayesian optimizer
(Section~\ref{sec:bayesian-optimization}), the acquisition function
is maximized to select a candidate.  However, the result is a value in
a continuous input space, while the hardware design space is discrete.
So Polaris instead enumerates the entire discrete hardware design space of
$8 \times 32 \times 32$ designs defined in Table~\ref{tab:design-space},
and then Polaris selects the candidate that maximizes the value of
the acquisition function.  Figure~\ref{fig:polaris} shows this process;
\FilledCircle{H1} shows with shapes the candidates in the hardware design
space, and \FilledCircle{H2} shows the candidates being assessed by the
acquisition function.  The candidate that maximizes the acquisition function
is shown with a filled circle, and it is fed as input to the software
optimizer.  The hardware optimization process is repeated for $n$ iterations.

\subsection{Layerwise Software Optimizer}

Given a hardware candidate, the layerwise software optimizer finds optimized
software mappings layer-by-layer.  The process is similar to that of the
hardware optimizer.

The first step is to select a software candidate.  Because the software design
space defined in Table~\ref{tab:design-space} is much larger than the hardware
design space, it is infeasible to evaluate the acquisition function for every
software design.  To reduce the number of designs without deteriorating their
quality, we enforce three reasonable constraints:  (1) the designs must be
implementable on the selected hardware candidate, (2) the spatially unrolled
dimensions---the C and K dimensions for Gemmini---should maximize the
utilization of the hardware, and (3) the tiling factors should evenly divide the
shape of the layer so that there are no extraneous edge cases that increase the
tail latency when running the layer.  Even after applying these constraints, the
software design space can still contain millions of points.  Thus, Polaris
selects a software candidate as follows:  It randomly draws using a Sobol
sequence 10,000 samples from the large, constrained software space, and it then
assesses each of the candidates with the acquisition function, selecting the
software candidate that maximizes the acquisition function.  This process is
shown in Figure~\ref{fig:polaris} \FilledCircle{S1} and \FilledCircle{S2}.  The
software candidate that is selected is shown with a heart, and the hardware
candidate selected by the hardware optimizer is still shown with a circle.

Once the HW/SW candidate is selected, it is evaluated on an RTL simulator,
FireSim~\cite{karandikar2018firesim}, as shown in \FilledCircle{S3}, and
Starlight is trained with the new evaluation, as shown in \FilledCircle{S4}.
The software optimization process repeats for {\it m} iterations.

\section{Methodology}
\label{sec:methodology}

This section presents our evaluation methodology.  We first describe the
DL models that we use to evaluate Starlight and Polaris.  We then describe
our methodology for evaluating each tool.

\subsection{DL Models}

We first train Starlight and Starlight-Low to predict the energy-delay
product (EDP) of executing individual layers from the following set of
diverse DL models.  We then use Polaris to perform HW/SW co-design for
each of these models independently.

\begin{itemize}
    \item U-Net~\cite{ronneberger2015unet} is a large convolutional neural
          network (CNN) used for biomedical image segmentation.

    \item ResNet-50~\cite{he2016deep} is a CNN used for image classification.

    \item BERT~\cite{devlin2019bert} is a transformer used for natural language
          processing.

    \item RetinaNet~\cite{lin2017focal} is CNN that adds to ResNet-50
          a feature pyramid network, classification head, and regression
          head.  We only evaluate these added layers.

\end{itemize}

\subsection{Starlight Methodology}

We measure the accuracy of Starlight and Starlight-Low using Spearman
rank correlation, $\rho$~\cite{fieller1957tests}, which compares the
relative ordering---as opposed to the precise value---of the predicted and
ground truth measurements.  $\rho$ ranges from -1 to +1, where -1 means
the relative orders are exactly reversed and +1 means the relative orders
are identical.  Because Starlight is used as the surrogate model of Polaris,
it need not predict the absolute performance value with high accuracy;  it is
sufficient for it to have high positive $\rho$.\footnote{For the sake of
completion, we also measure the typical accuracy metric---coorelation
coefficient---to be 97\%.}

Unless otherwise specified, we use 80\% of the datasets described in
Section~\ref{sec:starlight-dataset} for training and 20\% for testing.

\subsection{Polaris Methodology}
We compare Polaris against three baselines.

First, Offline Random draws random samples from the hardware and software
design spaces and evaluates them on Starlight.  The configuration that
minimizes EDP as predicted by Starlight is evaluated using RTL simulation.
By using Starlight for evaluation, Offline Random allows us to make a direct
comparison between offline optimization---i.e., optimizing a proxy model
without evaluating intermediate candidates using RTL simulation---and online
optimization---i.e., performing optimization by evaluating intermediate
candidates using RTL simulation.  We use random optimization instead of
optimization strategies such as stochastic gradient descent because the
latter optimizes the proxy model in a continuous space, and we find that
``snapping'' the results to an implementable design in the discrete design
space yields designs that are three orders of magnitude less performant
than those produced by random optimization.

Second, DOSA~\cite{hong2023dosa} is a state-of-the-art HW/SW co-design
tool that uses the same evaluation methodology as Polaris.  DOSA performs
stochastic gradient descent on a data-driven proxy model to find a HW/SW
configuration that minimizes EDP as predicted by the proxy model.  The
resulting design is evaluated using RTL simulation.  DOSA performs offline
optimization on a smaller design space than Polaris: It does not explore the
spatial array dimensions, and it only explores three possible loop orders.
We address these limitations in Section~\ref{sec:evaluation-polaris-sw}.

Third, Spotlight~\cite{sakhuja2023leveraging} is a state-of-the-art HW/SW
co-design tool that performs a feature transformation to improve the
sample-efficiency of a vanilla BO framework.  We adapt the methodology
used for Spotlight to evaluate candidates using RTL simulation in Polaris'
design space.  Spotlight performs online optimization.

We evaluate our tools in two design scenarios.  First, we perform HW/SW
co-design as described previously.  But since DOSA critically does not
include the spatial array dimensions in its design space, we evaluate the
baselines in a second design scenario:  software DSE.  When performing
software DSE, we use the DLA designs found by DOSA and only perform
layerwise software optimization.

When performing HW/SW co-design, Spotlight and Polaris run for $n=8$, $m=6$
iterations.  For fairness, Offline Random draws $8 \times 6 \times 10000
= 480000$ samples per layer from the HW/SW co-design space (recall that
Polaris evaluates 10,000 samples on the acquisition function per software
iteration).  When performing software DSE, Spotlight and Polaris run for
$m=20$ iterations, and Offline Random draws $20 \times 10000 = 200000$
samples per layer from the software design space.  Polaris and Spotlight
both run for three independent trials, and the median, minimum, and maximum of
the trials are reported.

\begin{figure}[t]
    \centering
    \includegraphics[width=0.6\linewidth]{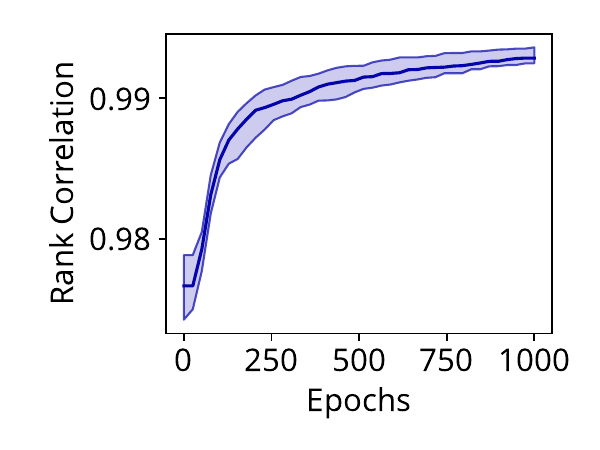}

    \caption{Starlight predicts EDP measured by RTL simulation, with
    Spearman rank correlation ($\rho$) of 0.99 after 1000 epochs of training.
    Across 10 independent trials, Starlight consistently achieves a median
    $\rho$, shown with the solid line, of greater than 0.98 within 100
    epochs. Furthermore, the narrowness of the shaded region, which denotes
    the cumulative minimum and maximum $\rho$ across the 10 trials, shows
    that Starlight trains accurately irrespective of the specific partition
    of training data that is used.  Higher is better.}

    \label{fig:starlight-convergence}

\end{figure}

\begin{figure}
    \centering
    \begin{subfigure}[b]{0.45\linewidth}
        \centering
        \includegraphics[width=\textwidth]{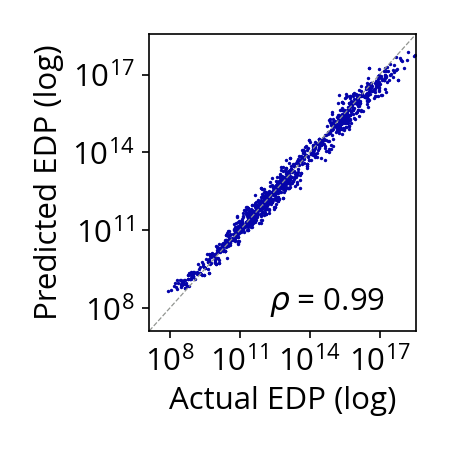}

        \caption{Starlight}

        \label{fig:starlight-firesim-corr}

    \end{subfigure}
    \hspace{0.03in}
    \begin{subfigure}[b]{0.45\linewidth}
        \centering

        \includegraphics[width=\textwidth]{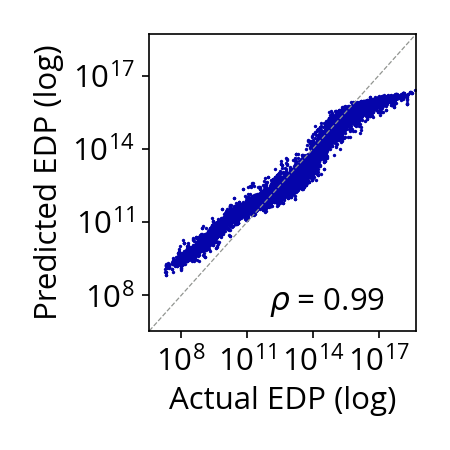}

        \caption{Starlight-Low}

        \label{fig:starlight-timeloop-corr}

    \end{subfigure}

    \caption{Accuracy and Spearman rank correlation ($\rho$) of
    the actual EDP and the predicted EDP for (a) Starlight and (b)
    Starlight-Low. Perfect accuracy is $y=x$ and $\rho=1$.}

\end{figure}

\section{Evaluation}
\label{sec:evaluation}

\ignore{
This section first evaluates Starlight and then Polaris, followed by a
discussion of our results.
}

\subsection{Starlight}
\label{sec:evaluation-starlight}

This section evaluates Starlight's accuracy and robustness.

\subsubsection{Accuracy}

Figure~\ref{fig:starlight-convergence} presents $\rho$ during training.
We perform 10 independent trials of training and plot the median (denoted
by the central line) and cumulative minimum and maximum (denoted by the
shaded region).  Starlight achieves $\rho = 0.99$ after 1000 epochs of
training (2 minutes of training time on a consumer-grade CPU), and it
consistently achieves $\rho \ge 0.98$ after just 100 trials (13 seconds
of training time on a consumer-grade CPU).  Because the shaded region is narrow
(standard deviation of $3.9 \times 10^{-4}$), we conclude that Starlight is not
sensitive to the specific partition of the training data that is used.  Note
that the initial accuracy is already high, demonstrating that transfer learning
provides significant benefit.  We further investigate this claim in
Section~\ref{sec:evaluation-robustness}.

Figures~\ref{fig:starlight-firesim-corr}
and~\ref{fig:starlight-timeloop-corr} show the accuracy and $\rho$ for
Starlight and Starlight-Low, respectively.  The X axis shows the ground
truth EDP measured by either FireSim for Starlight or by Timeloop for
Starlight-Low, and the Y axis shows the predicted EDP. Each dot represents a
sample from the test set.  If a sample is predicted with perfect accuracy,
it aligns with $y=x$.  Both Starlight and Starlight-Low achieve high
accuracy---as is indicated by the average distance across samples from
$y=x$---and a $\rho$ of 0.99.

\textbf{Key Takeaway}: Starlight achieves high accuracy when predicting
EDP as measured by RTL simulation.

\begin{figure}[t]
    \centering
    \includegraphics[width=\linewidth]{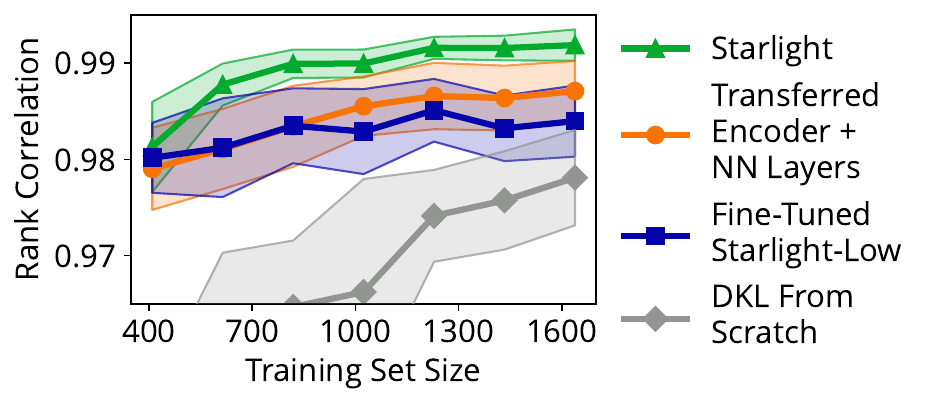}

    \caption{$\rho$ versus the FireSim training set size.  We evaluate four
    model architectures: (1) Starlight, (2) a neural network that leverages
    transfer learning, (3) a simple fine-tuning of Starlight-Low, and (4)
    a DKL trained from scratch.  The solid line indicates the mean of ten
    trials, and the shaded region indicates one standard deviation. Starlight
    consistently achieves the highest $\rho$ and is more resilient to the
    training set size and partition than other models. Higher is better.}

    \label{fig:transfer-learning-results}
\end{figure}

\subsubsection{Robustness}
\label{sec:evaluation-robustness}

Starlight achieves higher accuracy than Starlight-Low on
their respective datasets.  By comparing against three other performance
estimation approaches, we show, that Starlight's high accuracy can be
attributed to the use of both transfer learning and deep kernel learning
(DKL).  First, we compare against a model based on DKL with the same
architecture as Starlight but that is trained from scratch (DKL From
Scratch).  We then compare against a model that employs transfer learning
but trains a neural network predictor rather than a model based on DKL
(Transferred Encoder + NN Layers).  Finally, we compare against a simple
fine-tuning of the source model, Starlight-Low, that is trained without
the use of transfer learning (Fine-Tuned Starlight-Low).

Each model is trained with a range of training set sizes, and the training
process and the partitioning of the training set are repeated for 10
independent trials.  Figure~\ref{fig:transfer-learning-results} shows
the results.  The X axis shows the number of samples in the training set,
and the Y axis shows $\rho$ when each model predicts EDP of the test set.
The solid line indicates the mean of the trials, and the shaded region
indicates 1 standard deviation across the trials.

Starlight consistently achieves the highest $\rho$ out of the evaluated
models, irrespective of training set size.  Furthermore, Starlight achieves
the smallest standard deviation across trials, indicating that it is the
most robust of the evaluated models.

When trained on the full training set, DKL From Scratch achieves a mean
of $\rho = 0.973$.  Although this is high, it is significantly lower than
the other models evaluated, and the accuracy quickly deteriorates if the
training set size is reduced.  Overall, DKL From Scratch consistently
achieves the lowest accuracy.  However, Starlight also employs a model
based on DKL, so we conclude that DKL requires a large amount of data,
but it can be robust and achieve high accuracy.

To isolate the effects of transfer learning, we also compare against
Transferred Encoder + NN Layers.  This model achieves high accuracy,
but Starlight consistently achieves higher accuracy, indicating that
transfer learning is beneficial.  But DKL gives Starlight an edge over
other approaches.

Finally, to further validate our use of transfer learning, We compare
against Fine-Tuned Starlight-Low.  This model achieves high accuracy, but
Starlight and Transferred Encoder + NN Layers both consistently achieve
higher accuracy.

\textbf{Key Takeaway}:  Both DKL and transfer learning are instrumental
to Starlight's high accuracy and robustness.

\begin{figure*}[t]
    \centering
    \includegraphics[width=\linewidth]{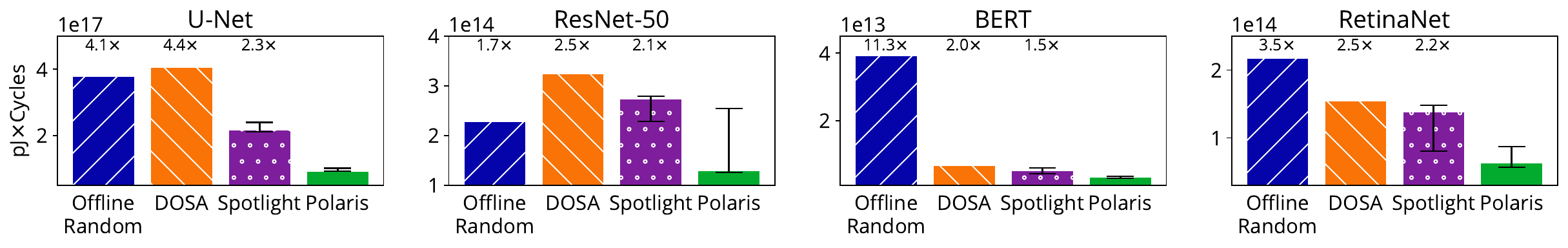}

    \caption{We compare the best designs produced by Offline Random, DOSA,
    Spotlight, and Polaris when performing HW/SW co-design to minimize EDP.
    Lower is better.}

    \label{fig:polaris-comparison-hw}
\end{figure*}

\begin{figure*}[t]
    \centering
    \includegraphics[width=\linewidth]{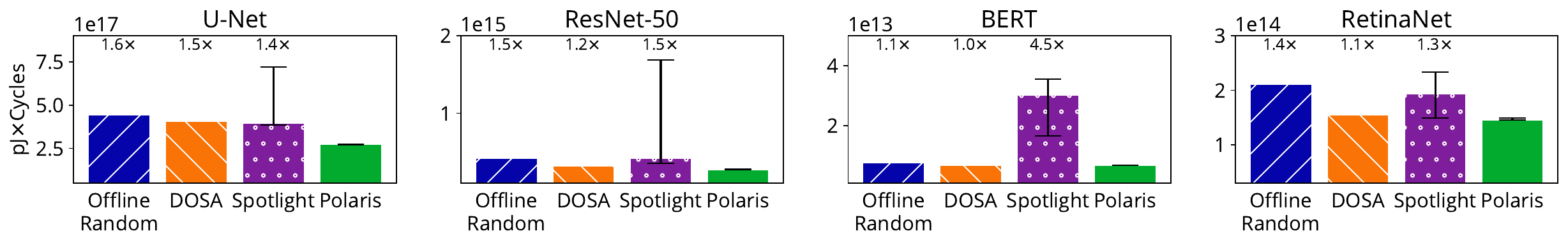}

    \caption{We compare the best software mappings produced by Offline,
    DOSA, Spotlight, and Polaris when performing software DSE to minimize
    EDP on the DLA design selected by DOSA.  Lower is better.}

    \label{fig:polaris-comparison-sw}
\end{figure*}

\subsection{Polaris}
\label{sec:evaluation-polaris}

In this section, we first demonstrate Polaris' advantage over prior work
when performing HW/SW co-design and software DSE.  We then compare the
behavior of our online methods: Spotlight and Polaris.

\subsubsection{HW/SW Co-Design}

Figure~\ref{fig:polaris-comparison-hw} compares the EDP of the designs
produced by Offline Random, DOSA, Spotlight, and Polaris when performing
HW/SW co-design.  The bars indicate the median of 3 independent trials,
and the error bars indicate the minimum and maximum of the trials.

In the median, Polaris consistently produces designs with the lowest EDP,
and Spotlight always produces designs that achieve lower EDP than those
produced by DOSA.  Part of this success can be attributed to the selection of
spatial array size, which is a design parameter that greatly affects EDP and
that is not explored by DOSA. In particular, the 32$\times$32 spatial array
is almost always selected by Polaris and Spotlight because it reduces EDP.
In the outlying maximum trial of ResNet-50, Polaris does not select a
32$\times$32 spatial array.  Similarly, DOSA always uses a 16$\times$16 spatial
array, so it achieves higher EDP than Polaris and Spotlight.  However, the
spatial array size is not the sole reason for Polaris' success; Offline
Random always selects a 32$\times$32 spatial array, but it is unable to
select other commensurate design parameter values, so its designs always
result in higher EDP than both Spotlight and Polaris.

\textbf{Key Takeaway}:  The online methods, Polaris and Spotlight,
consistently produce designs with lower EDP than the offline methods when
performing HW/SW co-design, and Polaris consistently produces designs with
the lowest median EDP.

\subsubsection{Software DSE}
\label{sec:evaluation-polaris-sw}

Because the DLA architecture significantly affects the final EDP found, we
select a fixed DLA design---specifically, the DLA design selected by DOSA---and
only perform software DSE to produce software mappings.
Figure~\ref{fig:polaris-comparison-sw} summarizes these results.

We again find that Polaris consistently produces designs that achieve the
lowest EDP, but its improvement over the baselines is smaller.  Furthermore,
the EDP achieved when Polaris performs software DSE is consistently higher
than the EDP achieved when Polaris performs HW/SW co-design.  These results
corroborate prior work~\cite{parashar2019timeloop,shao2019simba} that
highlights the importance of performing HW/SW co-design.

We also find that Spotlight no longer consistently produces software mappings
with lower EDP than those produced by DOSA, and the variance across trials is
significantly higher.  We hypothesize that Spotlight, which is trained from
scratch on each trial, struggles to learn the characteristics of the design
space when it has fewer degrees of freedom to learn from.  On the other hand,
Polaris's proxy model, Starlight, is already trained on the design space and can
navigate it effectively from the start.

\textbf{Key Takeaway}:  Polaris consistently produces software mappings
that achieve lower EDP than the baselines when performing software DSE, and
in general it is important to perform HW/SW co-design when designing DLAs.

\begin{figure*}[t]
    \centering
    \includegraphics[width=\linewidth]{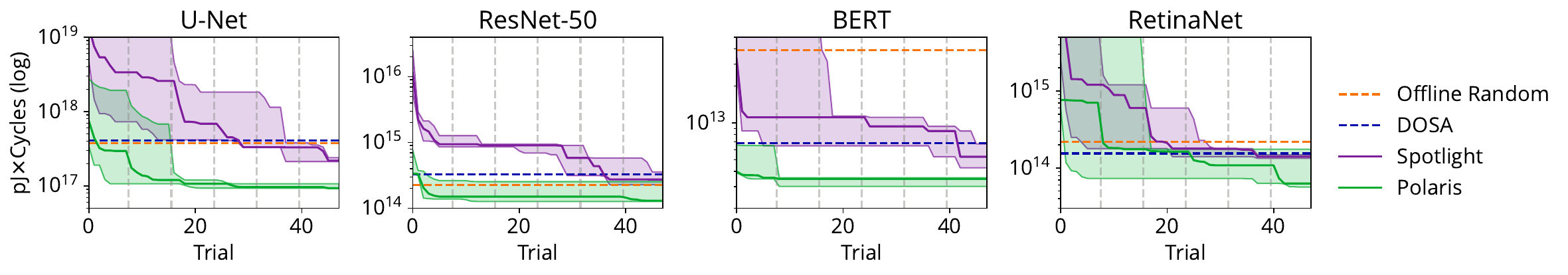}

    \caption{The behavior Polaris and Spotlight when performing HW/SW
    co-design.  Each segment demarcated by a gray dashed line is a single
    hardware candidate, and the solid line indicates the cumulative minimum
    EDP found thus far. Lower is better.}

    \label{fig:polaris-convergence-hw}
\end{figure*}

\subsubsection{Online Optimization Behavior}

Figure~\ref{fig:polaris-convergence-hw} summarizes our investigation
into the behavior of online optimization.  The X axis shows the overall
iteration of the hardware and software optimizers, and the gray dashed lines
demarcate the 8 hardware candidates that are evaluated.  The Y axis shows
the cumulative minimum EDP that has been achieved thus far.  For Polaris
and Spotlight, the solid line indicates the median of 3 independent trials,
and the shaded region indicates the minimum and maximum across the trials.

For U-Net, ResNet-50, and BERT, Polaris' optimization quickly converge,
whereas Spotlight's optimization is more segmented.  This behavior is
unsurprising because Polaris is trained on a dataset of RTL simulations,
so it begins the optimization with a noteworthy head-start.  Spotlight
begins with 3 uniformly random samples to seed the surrogate model, and it
continues to learn the shape of the design space to reduce its achieved EDP.
If given more iterations, Spotlight may eventually produce results similar
to Polaris's, but we explain in Section~\ref{sec:discussion} why Polaris is
still a better choice for HW/SW co-design when the evaluation method is slow.

Finally, across all models and for both Polaris and Spotlight, the biggest
changes in EDP occur when a new hardware candidate is selected---i.e., at
the grey dashed lines.  Thus, we see that the choice of hardware candidate
plays a significant role in the final achievable EDP, but the software
mappings must properly optimized for the hardware.

We perform this same analysis for software DSE, but we omit the results
for brevity because they are similar to that of HW/SW co-design.

\textbf{Key Takeaway}:  Polaris and Spotlight both find designs with low
EDP when performing HW/SW co-design, but Polaris is more sample-efficient.
Furthermore, the choice of hardware candidate plays a significant role in
the final achievable EDP.

\subsection{Discussion}
\label{sec:discussion}

\begin{table}
    \centering

    \begin{tabular}{|l|l|l|l|l|}
        \hline
        & \multicolumn{2}{|c|}{\textbf{HW/SW Co-Design}} & \multicolumn{2}{|c|}{\textbf{SW DSE}} \\
        \hline
        \hline
        \textbf{Model} & \textbf{Spotlight} & \textbf{Polaris} & \textbf{Spotlight} & \textbf{Polaris} \\
        \hline
        U-Net & 9.19h & 0.35h & 2.28h & 0.31h \\
        \hline
        ResNet-50 & 1.70h & 0.29h & - & 0.44h \\
        \hline
        BERT & 0.98h & 0.10h & - & 0.39h \\
        \hline
        RetinaNet & 1.37h & 1.60h & 0.41h & 0.92h \\
        \hline
    \end{tabular}

    \caption{The wall-clock time for each online method---when performing
    HW/SW co-design and software DSE---to produce designs that achieve lower EDP
    than the designs found by DOSA.  Lower is better.}

    \label{tab:wall-clock-time}

\end{table}

Offline Random and DOSA both perform offline optimization,
while Spotlight and Polaris both perform online optimization.
Figures~\ref{fig:polaris-comparison-hw} and~\ref{fig:polaris-comparison-sw}
both clearly illustrate the benefit of performing online optimization:
The online methods can produce better designs than the offline methods.

Another important metric to compare the quality of these tools is the amount of
wall-clock time it takes for the online methods to outperform the offline
methods.  Table~\ref{tab:wall-clock-time} presents these results.  Within a
maximum of 1.6 hours and an average of 35 minutes, Polaris always produces
designs that outperform those produced by DOSA in 6 hours, and the remainder
of the time is spent exploring designs that achieve even lower EDP.  And because
intermediate designs are evaluated using RTL simulation, both Polaris and
Spotlight continuously learn more about the design space and will likely
continue to reduce EDP with each iteration.

Polaris is generally faster than Spotlight at producing designs that
outperform those produced by DOSA.  There are two reasons for this.
First, Polaris is warmed up with the dataset of RTL simulations, so it
is able to quickly find designs that achieve low EDP.  Second, the RTL
simulations for Polaris' candidate designs have shorter wall-clock time
than that of Spotlight's candidates designs because the wall-clock time of
RTL simulation is correlated with the delay of the design being simulated,
and Spotlight's designs typically have higher delay.

Of course, Polaris requires a dataset of Timeloop evaluations and RTL
simulations to be collected beforehand, which incurs an 8-hour, one-time cost
that Spotlight does not incur.  However, in practice, Polaris will be run
multiple times over the course of the DLA development cycle, so the cost of
collecting the dataset is amortized.  Over time, Polaris provides significantly
higher sample-efficiency than Spotlight, so it is the better choice for HW/SW
co-design when the evaluation method is slow.

Polaris achieves its high sample-efficiency by spending more wall-clock
time than Spotlight to select candidates to evaluate using RTL simulation.
But in situations where the evaluation method is fast, such as early in the
development cycle when designs are evaluated using an analytical model,
approaches like Spotlight may be better suited because they can evaluate
a larger number of candidates.

\section{Conclusions}
\label{sec:conclusion}

In this paper, we have shown that transfer learning can be effectively
employed to transfer knowledge from a low-fidelity performance model to
a high-fidelity performance model.  In particular, we have shown that
a data-driven model that has been trained using a fast analytical model
can reduce the number of slow evaluations needed to train a high-fidelity
data-driven model.  Our resulting data-driven model can be queried faster
than an analytical model and can achieve 99\% accuracy when predicting the
EDP of a DLA.  Moreover, Starlight is trained with 61\% fewer high-fidelity
evaluations and achieves higher accuracy than DOSA's state-of-the-art
data-driven model~\cite{hong2023dosa}.

We have also used Starlight as a key component in Polaris, which is the
first DSE tool that performs RTL simulation in the optimization loop.
Polaris produces designs that reduce the energy-delay product by 2.7$\times$
over DOSA and by 5.15$\times$ over an ablated version of Polaris that does
not perform RTL simulation in the optimization loop.

The methodology that we have presented in this paper may be applicable to
other areas of hardware design in general where there is close similarity
between low-fidelity evaluation methods and high-fidelity evaluation
methods and where high-fidelity evaluation methods are slow.  Irrespective
of its broader applicability, our methodology indicates the importance
of exploiting properties of the problem---e.g., the transferability of
knowledge between low and high fidelity performance models---to design
customized design space exploration tools.

\ignore{
\section*{Acknowledgments}

We thank Kevin Swersky for his discussions on transfer learning.  This work
was funded in part by NSF Grant CCF-1823546, a gift from Intel Corporation
through the NSF/Intel Partnership on Foundational Microarchitecture Research,
an NXP fellowship, and a gift from Arm, Inc.
}


\bibliographystyle{IEEEtranS}
\bibliography{refs}

\begin{thebibliography}{10}
\providecommand{\url}[1]{#1}
\csname url@samestyle\endcsname
\providecommand{\newblock}{\relax}
\providecommand{\bibinfo}[2]{#2}
\providecommand{\BIBentrySTDinterwordspacing}{\spaceskip=0pt\relax}
\providecommand{\BIBentryALTinterwordstretchfactor}{4}
\providecommand{\BIBentryALTinterwordspacing}{\spaceskip=\fontdimen2\font plus
\BIBentryALTinterwordstretchfactor\fontdimen3\font minus
  \fontdimen4\font\relax}
\providecommand{\BIBforeignlanguage}[2]{{%
\expandafter\ifx\csname l@#1\endcsname\relax
\typeout{** WARNING: IEEEtranS.bst: No hyphenation pattern has been}%
\typeout{** loaded for the language `#1'. Using the pattern for}%
\typeout{** the default language instead.}%
\else
\language=\csname l@#1\endcsname
\fi
#2}}
\providecommand{\BIBdecl}{\relax}
\BIBdecl

\bibitem{abts2022challenges}
D.~Abts, I.~Ahmed, A.~Bitar, M.~Boyd, J.~Kim, G.~Kimmell, and A.~Ling,
  ``Challenges/{{Opportunities}} to {{Enable Dependable Scale-out System}} with
  {{Groq Deterministic Tensor-Streaming Processors}},'' in \emph{Dependable
  {{Systems}} and {{Networks}} ({{DSN-S}})}, Jun. 2022.

\bibitem{bai2021boomexplorer}
C.~Bai, Q.~Sun, J.~Zhai, Y.~Ma, B.~Yu, and M.~D. Wong, ``{{BOOM-Explorer}}:
  {{RISC-V BOOM Microarchitecture Design Space Exploration Framework}},'' in
  \emph{International {{Conference On Computer-Aided Design}} ({{ICCAD}})},
  Nov. 2021.

\bibitem{bai2023transfer}
T.~Bai, Y.~Li, Y.~Shen, X.~Zhang, W.~Zhang, and B.~Cui, ``Transfer {{Learning}}
  for {{Bayesian Optimization}}: {{A Survey}},'' \emph{arXiv}, Feb. 2023.

\bibitem{bank2023autoencoders}
D.~Bank, N.~Koenigstein, and R.~Giryes, ``Autoencoders,'' in \emph{Data
  {{Mining}} and {{Knowledge Discovery Handbook}}}, L.~Rokach, O.~Maimon, and
  E.~Shmueli, Eds., 2023.

\bibitem{bischl2023hyperparameter}
B.~Bischl, M.~Binder, M.~Lang, T.~Pielok, J.~Richter, S.~Coors, J.~Thomas,
  T.~Ullmann, M.~Becker, A.-L. Boulesteix, D.~Deng, and M.~Lindauer,
  ``Hyperparameter {{Optimization}}: {{Foundations}}, {{Algorithms}}, {{Best
  Practices}}, and {{Open Challenges}},'' \emph{WIREs Data Mining and Knowledge
  Discovery}, no.~2, 2023.

\bibitem{chen2017eyeriss}
Y.-H. Chen, T.~Krishna, J.~S. Emer, and V.~Sze, ``Eyeriss: {{An
  Energy-Efficient Reconfigurable Accelerator}} for {{Deep Convolutional Neural
  Networks}},'' \emph{Solid-State Circuits}, no.~1, Jan. 2017.

\bibitem{chen2019eyeriss}
Y.-H. Chen, T.-J. Yang, J.~Emer, and V.~Sze, ``Eyeriss v2: {{A Flexible
  Accelerator}} for {{Emerging Deep Neural Networks}} on {{Mobile Devices}},''
  \emph{Emerging and Selected Topics in Circuits and Systems}, no.~2, Jun.
  2019.

\bibitem{dave2019dmazerunner}
S.~Dave, Y.~Kim, S.~Avancha, K.~Lee, and A.~Shrivastava, ``{{dMazeRunner}}:
  {{Executing Perfectly Nested Loops}} on {{Dataflow Accelerators}},''
  \emph{Transactions on Embedded Computing Systems}, no.~5s, Oct. 2019.

\bibitem{devlin2019bert}
J.~Devlin, M.-W. Chang, K.~Lee, and K.~Toutanova, ``{{BERT}}: {{Pre-training}}
  of {{Deep Bidirectional Transformers}} for {{Language Understanding}},''
  \emph{arXiv}, May 2019.

\bibitem{dhilleswararao2022efficient}
P.~Dhilleswararao, S.~Boppu, M.~S. Manikandan, and L.~R. Cenkeramaddi,
  ``Efficient {{Hardware Architectures}} for {{Accelerating Deep Neural
  Networks}}: {{Survey}},'' \emph{IEEE Access}, 2022.

\bibitem{dong2021survey}
S.~Dong, P.~Wang, and K.~Abbas, ``A {{Survey}} on {{Deep Learning}} and {{Its
  Applications}},'' \emph{Computer Science Review}, 2021.

\bibitem{emani2021accelerating}
M.~Emani, V.~Vishwanath, C.~Adams, M.~E. Papka, R.~Stevens, L.~Florescu,
  S.~Jairath, W.~Liu, T.~Nama, and A.~Sujeeth, ``Accelerating {{Scientific
  Applications With SambaNova Reconfigurable Dataflow Architecture}},''
  \emph{Computing in Science \& Engineering}, no.~2, Mar. 2021.

\bibitem{esmaeilzadeh2023opensource}
H.~Esmaeilzadeh, S.~Ghodrati, A.~B. Kahng, J.~K. Kim, S.~Kinzer, S.~Kundu,
  R.~Mahapatra, S.~D. Manasi, S.~Sapatnekar, Z.~Wang, and Z.~Zeng, ``An
  {{Open-Source ML-Based Full-Stack Optimization Framework}} for {{Machine
  Learning Accelerators}},'' \emph{arXiv}, Aug. 2023.

\bibitem{esmaeilzadeh2022physically}
H.~Esmaeilzadeh, S.~Ghodrati, A.~B. Kahng, J.~K. Kim, S.~Kinzer, S.~Kundu,
  R.~Mahapatra, S.~D. Manasi, S.~S. Sapatnekar, Z.~Wang, and Z.~Zeng,
  ``Physically {{Accurate Learning-Based Performance Prediction}} of
  {{Hardware-Accelerated ML Algorithms}},'' in \emph{Workshop on {{Machine
  Learning}} for {{CAD}} ({{MLCAD}})}, Sep. 2022.

\bibitem{ferianc2021improving}
M.~Ferianc, H.~Fan, D.~Manocha, H.~Zhou, S.~Liu, X.~Niu, and W.~Luk,
  ``Improving {{Performance Estimation}} for {{Design Space Exploration}} for
  {{Convolutional Neural Network Accelerators}},'' \emph{Electronics}, no.~4,
  Jan. 2021.

\bibitem{feurer2022practical}
M.~Feurer, B.~Letham, F.~Hutter, and E.~Bakshy, ``Practical {{Transfer
  Learning}} for {{Bayesian Optimization}},'' \emph{arXiv}, Oct. 2022.

\bibitem{fieller1957tests}
E.~C. Fieller, H.~O. Hartley, and E.~S. Pearson, ``Tests for {{Rank Correlation
  Coefficients}}, {{I}},'' \emph{Biometrika}, no. 3-4, Dec. 1957.

\bibitem{forrester2007multifidelity}
A.~I. Forrester, A.~S{\'o}bester, and A.~J. Keane, ``Multi-fidelity
  {{Optimization}} via {{Surrogate Modelling}},'' \emph{Royal Society A:
  Mathematical, Physical and Engineering Sciences}, no. 2088, Oct. 2007.

\bibitem{genc2021gemmini}
H.~Genc, S.~Kim, A.~Amid, A.~{Haj-Ali}, V.~Iyer, P.~Prakash, J.~Zhao, D.~Grubb,
  H.~Liew, H.~Mao, A.~Ou, C.~Schmidt, S.~Steffl, J.~Wright, I.~Stoica,
  J.~{Ragan-Kelley}, K.~Asanovic, B.~Nikolic, and Y.~S. Shao, ``Gemmini:
  {{Enabling Systematic Deep-Learning Architecture Evaluation}} via
  {{Full-Stack Integration}},'' in \emph{Design {{Automation Conference}}
  ({{DAC}})}, Dec. 2021.

\bibitem{gomez-bombarelli2018automatic}
R.~{G{\'o}mez-Bombarelli}, J.~N. Wei, D.~Duvenaud, J.~M.
  {Hern{\'a}ndez-Lobato}, B.~{S{\'a}nchez-Lengeling}, D.~Sheberla,
  J.~{Aguilera-Iparraguirre}, T.~D. Hirzel, R.~P. Adams, and A.~{Aspuru-Guzik},
  ``Automatic {{Chemical Design Using}} a {{Data-Driven Continuous
  Representation}} of {{Molecules}},'' \emph{ACS Central Science}, no.~2, Feb.
  2018.

\bibitem{he2016deep}
K.~He, X.~Zhang, S.~Ren, and J.~Sun, ``Deep {{Residual Learning}} for {{Image
  Recognition}},'' in \emph{Computer {{Vision}} and {{Pattern Recognition}}
  ({{CVPR}})}, 2016.

\bibitem{hong2023dosa}
C.~Hong, Q.~Huang, G.~Dinh, M.~Subedar, and Y.~S. Shao, ``{{DOSA}}:
  {{Differentiable Model-Based One-Loop Search}} for {{DNN Accelerators}},'' in
  \emph{Microarchitecture ({{MICRO}})}, Dec. 2023.

\bibitem{huang2022learning}
Q.~Huang, C.~Hong, J.~Wawrzynek, M.~Subedar, and Y.~S. Shao, ``Learning {{A
  Continuous}} and {{Reconstructible Latent Space}} for {{Hardware Accelerator
  Design}},'' in \emph{International {{Symposium}} on {{Performance Analysis}}
  of {{Systems}} and {{Software}} ({{ISPASS}})}, May 2022.

\bibitem{jouppi2021ten}
N.~P. Jouppi, D.~Hyun~Yoon, M.~Ashcraft, M.~Gottscho, T.~B. Jablin, G.~Kurian,
  J.~Laudon, S.~Li, P.~Ma, X.~Ma, T.~Norrie, N.~Patil, S.~Prasad, C.~Young,
  Z.~Zhou, and D.~Patterson, ``Ten {{Lessons From Three Generations Shaped
  Google}}'s {{TPUv4i}} : {{Industrial Product}},'' in \emph{International
  {{Symposium}} on {{Computer Architecture}} ({{ISCA}})}, Jun. 2021.

\bibitem{jouppi2017indatacenter}
N.~P. Jouppi, C.~Young, N.~Patil, D.~Patterson, G.~Agrawal, R.~Bajwa, S.~Bates,
  S.~Bhatia, N.~Boden, A.~Borchers, R.~Boyle, P.-l. Cantin, C.~Chao, C.~Clark,
  J.~Coriell, M.~Daley, M.~Dau, J.~Dean, B.~Gelb, T.~V. Ghaemmaghami,
  R.~Gottipati, W.~Gulland, R.~Hagmann, C.~R. Ho, D.~Hogberg, J.~Hu, R.~Hundt,
  D.~Hurt, J.~Ibarz, A.~Jaffey, A.~Jaworski, A.~Kaplan, H.~Khaitan,
  D.~Killebrew, A.~Koch, N.~Kumar, S.~Lacy, J.~Laudon, J.~Law, D.~Le, C.~Leary,
  Z.~Liu, K.~Lucke, A.~Lundin, G.~MacKean, A.~Maggiore, M.~Mahony, K.~Miller,
  R.~Nagarajan, R.~Narayanaswami, R.~Ni, K.~Nix, T.~Norrie, M.~Omernick,
  N.~Penukonda, A.~Phelps, J.~Ross, M.~Ross, A.~Salek, E.~Samadiani, C.~Severn,
  G.~Sizikov, M.~Snelham, J.~Souter, D.~Steinberg, A.~Swing, M.~Tan,
  G.~Thorson, B.~Tian, H.~Toma, E.~Tuttle, V.~Vasudevan, R.~Walter, W.~Wang,
  E.~Wilcox, and D.~H. Yoon, ``In-{{Datacenter Performance Analysis}} of a
  {{Tensor Processing Unit}},'' in \emph{International {{Symposium}} on
  {{Computer Architecture}} ({{ISCA}})}, Jun. 2017.

\bibitem{kao2020confuciux}
S.-C. Kao, G.~Jeong, and T.~Krishna, ``{{ConfuciuX}}: {{Autonomous Hardware
  Resource Assignment}} for {{DNN Accelerators}} using {{Reinforcement
  Learning}},'' in \emph{Microarchitecture ({{MICRO}})}, Oct. 2020.

\bibitem{karandikar2018firesim}
S.~Karandikar, H.~Mao, D.~Kim, D.~Biancolin, A.~Amid, D.~Lee, N.~Pemberton,
  E.~Amaro, C.~Schmidt, A.~Chopra, Q.~Huang, K.~Kovacs, B.~Nikolic, R.~Katz,
  J.~Bachrach, and K.~Asanovi{\'c}, ``Firesim: {{FPGA-Accelerated Cycle-Exact
  Scale-Out System Simulation}} in the {{Public Cloud}},'' in
  \emph{International {{Symposium}} on {{Computer Architecture}} ({{ISCA}})},
  Jun. 2018.

\bibitem{kaufman2021learned}
S.~Kaufman, P.~Phothilimthana, Y.~Zhou, C.~Mendis, S.~Roy, A.~Sabne, and
  M.~Burrows, ``A {{Learned Performance Model}} for {{Tensor Processing
  Units}},'' in \emph{Machine {{Learning}} and {{Systems}}}, A.~Smola,
  A.~Dimakis, and I.~Stoica, Eds., 2021.

\bibitem{kim2023full}
S.~Kim, C.~Hooper, T.~Wattanawong, M.~Kang, R.~Yan, H.~Genc, G.~Dinh, Q.~Huang,
  K.~Keutzer, M.~W. Mahoney, Y.~S. Shao, and A.~Gholami, ``Full {{Stack
  Optimization}} of {{Transformer Inference}}: A {{Survey}},'' \emph{arXiv},
  Feb. 2023.

\bibitem{kingma2022autoencoding}
D.~P. Kingma and M.~Welling, ``Auto-{{Encoding Variational Bayes}},''
  \emph{arXiv}, Dec. 2022.

\bibitem{koeplinger2018spatial}
D.~Koeplinger, M.~Feldman, R.~Prabhakar, Y.~Zhang, S.~Hadjis, R.~Fiszel,
  T.~Zhao, L.~Nardi, A.~Pedram, C.~Kozyrakis, and K.~Olukotun, ``Spatial: A
  language and compiler for application accelerators,'' in \emph{Programming
  {{Language Design}} and {{Implementation}} ({{PLDI}})}, Jun. 2018.

\bibitem{krishnan2023archgym}
S.~Krishnan, A.~Yazdanbakhsh, S.~Prakash, J.~Jabbour, I.~Uchendu, S.~Ghosh,
  B.~Boroujerdian, D.~Richins, D.~Tripathy, A.~Faust, and V.~Janapa~Reddi,
  ``{{ArchGym}}: {{An Open-Source Gymnasium}} for {{Machine Learning Assisted
  Architecture Design}},'' in \emph{International {{Symposium}} on {{Computer
  Architecture}} ({{ISCA}})}, Jun. 2023.

\bibitem{kullback1951information}
S.~Kullback and R.~A. Leibler, ``On {{Information}} and {{Sufficiency}},''
  \emph{The Annals of Mathematical Statistics}, no.~1, 1951.

\bibitem{kumar2021datadriven}
A.~Kumar, A.~Yazdanbakhsh, M.~Hashemi, K.~Swersky, and S.~Levine,
  ``Data-{{Driven Offline Optimization}} for {{Architecting Hardware
  Accelerators}},'' in \emph{International {{Conference}} on {{Learning
  Representations}} ({{ICLR}})}, Oct. 2021.

\bibitem{kwon2020maestro}
H.~Kwon, P.~Chatarasi, V.~Sarkar, T.~Krishna, M.~Pellauer, and A.~Parashar,
  ``{{MAESTRO}}: {{A Data-Centric Approach}} to {{Understand Reuse}},
  {{Performance}}, and {{Hardware Cost}} of {{DNN Mappings}},'' \emph{IEEE
  Micro}, no.~3, May 2020.

\bibitem{lattner2021mlir}
C.~Lattner, M.~Amini, U.~Bondhugula, A.~Cohen, A.~Davis, J.~Pienaar, R.~Riddle,
  T.~Shpeisman, N.~Vasilache, and O.~Zinenko, ``{{MLIR}}: {{Scaling Compiler
  Infrastructure}} for {{Domain Specific Computation}},'' in \emph{Code
  {{Generation}} and {{Optimization}} ({{CGO}})}, Feb. 2021.

\bibitem{lavely2022powering}
A.~Lavely, ``Powering {{Extreme-Scale HPC}} with {{Cerebras Wafer- Scale
  Accelerators}},'' Cerebras Systems, Inc, Tech. Rep., 2022.

\bibitem{li2023study}
Y.~L. Li, T.~G.~J. Rudner, and A.~G. Wilson, ``A {{Study}} of {{Bayesian Neural
  Network Surrogates}} for {{Bayesian Optimization}},'' \emph{arXiv}, May 2023.

\bibitem{lin2017focal}
T.-Y. Lin, P.~Goyal, R.~Girshick, K.~He, and P.~Dollar, ``Focal {{Loss}} for
  {{Dense Object Detection}},'' in \emph{International {{Conference}} on
  {{Computer Vision}} ({{ICCV}})}, 2017.

\bibitem{lin2021naas}
Y.~Lin, M.~Yang, and S.~Han, ``{{NAAS}}: {{Neural Accelerator Architecture
  Search}},'' in \emph{Design {{Automation Conference}} ({{DAC}})}, Dec. 2021.

\bibitem{maslej2023ai}
N.~Maslej, L.~Fattorini, E.~Brynjolfsson, J.~Etchemendy, K.~Ligett, T.~Lyons,
  J.~Manyika, H.~Ngo, V.~Parli, Y.~Shoham, R.~Wald, J.~Clark, and R.~Perrault,
  ``The {{AI Index}} 2023 {{Annual Report}},'' Institute for Human-Centered AI,
  Tech. Rep., Apr. 2023.

\bibitem{matern1986spatial}
B.~Mat{\'e}rn, \emph{Spatial {{Variation}}}, D.~Brillinger, S.~Fienberg,
  J.~Gani, J.~Hartigan, and K.~Krickeberg, Eds., 1986.

\bibitem{mei2021zigzag}
L.~Mei, P.~Houshmand, V.~Jain, S.~Giraldo, and M.~Verhelst, ``{{ZigZag}}:
  {{Enlarging Joint Architecture-Mapping Design Space Exploration}} for {{DNN
  Accelerators}},'' \emph{Transactions on Computers}, no.~8, Aug. 2021.

\bibitem{munoz-martinez2021stonne}
F.~{Mu{\~n}oz-Mart{\'i}nez}, J.~L. Abell{\'a}n, M.~E. Acacio, and T.~Krishna,
  ``{{STONNE}}: {{Enabling Cycle-Level Microarchitectural Simulation}} for
  {{DNN Inference Accelerators}},'' in \emph{International {{Symposium}} on
  {{Workload Characterization}} ({{IISWC}})}, Nov. 2021.

\bibitem{nardi2019practical}
L.~Nardi, D.~Koeplinger, and K.~Olukotun, ``Practical {{Design Space
  Exploration}},'' in \emph{Modeling, {{Analysis}}, and {{Simulation}} of
  {{Computer}} and {{Telecommunication Systems}} ({{MASCOTS}})}, Oct. 2019.

\bibitem{parashar2019timeloop}
A.~Parashar, P.~Raina, Y.~S. Shao, Y.-H. Chen, V.~A. Ying, A.~Mukkara,
  R.~Venkatesan, B.~Khailany, S.~W. Keckler, and J.~Emer, ``Timeloop: {{A
  Systematic Approach}} to {{DNN Accelerator Evaluation}},'' in
  \emph{International {{Symposium}} on {{Performance Analysis}} of {{Systems}}
  and {{Software}} ({{ISPASS}})}, Mar. 2019.

\bibitem{pham-quoc2021hardware}
C.~{Pham-Quoc}, X.-Q. Nguyen, and T.~N. Thinh, ``Hardware/{{Software
  Co-design}} for {{Convolutional Neural Networks Acceleration}}: {{A Survey}}
  and {{Open Issues}},'' in \emph{Context-{{Aware Systems}} and
  {{Applications}}}, P.~Cong~Vinh and A.~Rakib, Eds., 2021.

\bibitem{reagen2017case}
B.~Reagen, J.~M. {Hern{\'a}ndez-Lobato}, R.~Adolf, M.~Gelbart, P.~Whatmough,
  G.-Y. Wei, and D.~Brooks, ``A {{Case}} for {{Efficient Accelerator Design
  Space Exploration}} via {{Bayesian Optimization}},'' in \emph{International
  {{Symposium}} on {{Low Power Electronics}} and {{Design}} ({{ISLPED}})}, Jul.
  2017.

\bibitem{rezende2014stochastic}
D.~J. Rezende, S.~Mohamed, and D.~Wierstra, ``Stochastic {{Backpropagation}}
  and {{Approximate Inference}} in {{Deep Generative Models}},'' in
  \emph{International {{Conference}} on {{Machine Learning}} ({{ICML}})}, Jun.
  2014.

\bibitem{ronneberger2015unet}
O.~Ronneberger, P.~Fischer, and T.~Brox, ``U-{{Net}}: {{Convolutional
  Networks}} for {{Biomedical Image Segmentation}},'' in \emph{Medical {{Image
  Computing}} and {{Computer-Assisted Intervention}} ({{MICCAI}})}, N.~Navab,
  J.~Hornegger, W.~M. Wells, and A.~F. Frangi, Eds., 2015.

\bibitem{rumelhart1986learning}
D.~E. Rumelhart, G.~E. Hinton, and R.~J. Williams, ``Learning {{Internal
  Representations}} by {{Error Propagation}},'' in \emph{Explorations in the
  {{Microstructure}} of {{Cognition}}}, Jan. 1986.

\bibitem{sakhuja2023leveraging}
C.~Sakhuja, Z.~Shi, and C.~Lin, ``Leveraging {{Domain Information}} for the
  {{Efficient Automated Design}} of {{Deep Learning Accelerators}},'' in
  \emph{High-{{Performance Computer Architecture}} ({{HPCA}})}, Feb. 2023.

\bibitem{samajdar2023airchitect}
A.~Samajdar, J.~M. Joseph, and T.~Krishna, ``{{AIrchitect}}: {{Automating
  Hardware Architecture}} and {{Mapping Optimization}},'' in \emph{Design,
  {{Automation}} \& {{Test}} in {{Europe Conference}} \& {{Exhibition}}
  ({{DATE}})}, Apr. 2023.

\bibitem{samajdar2020systematic}
A.~Samajdar, J.~M. Joseph, Y.~Zhu, P.~Whatmough, M.~Mattina, and T.~Krishna,
  ``A {{Systematic Methodology}} for {{Characterizing Scalability}} of {{DNN
  Accelerators}} using {{SCALE-Sim}},'' in \emph{International {{Symposium}} on
  {{Performance Analysis}} of {{Systems}} and {{Software}} ({{ISPASS}})}, Aug.
  2020.

\bibitem{sekanina2021neural}
L.~Sekanina, ``Neural {{Architecture Search}} and {{Hardware Accelerator
  Co-Search}}: {{A Survey}},'' \emph{IEEE Access}, 2021.

\bibitem{seshadri2022evaluation}
K.~Seshadri, B.~Akin, J.~Laudon, R.~Narayanaswami, and A.~Yazdanbakhsh, ``An
  {{Evaluation}} of {{Edge TPU Accelerators}} for {{Convolutional Neural
  Networks}},'' in \emph{International {{Symposium}} on {{Workload
  Characterization}} ({{IISWC}})}, Nov. 2022.

\bibitem{shao2019simba}
Y.~S. Shao, J.~Clemons, R.~Venkatesan, B.~Zimmer, M.~Fojtik, N.~Jiang,
  B.~Keller, A.~Klinefelter, N.~Pinckney, P.~Raina, S.~G. Tell, Y.~Zhang, W.~J.
  Dally, J.~Emer, C.~T. Gray, B.~Khailany, and S.~W. Keckler, ``Simba:
  {{Scaling Deep-Learning Inference}} with {{Multi-Chip-Module-Based
  Architecture}},'' in \emph{Microarchitecture ({{MICRO}})}, Oct. 2019.

\bibitem{sipola2022artificial}
T.~Sipola, J.~Alatalo, T.~Kokkonen, and M.~Rantonen, ``Artificial
  {{Intelligence}} in the {{IoT Era}}: {{A Review}} of {{Edge AI Hardware}} and
  {{Software}},'' in \emph{Conference of {{Open Innovations Association}}
  ({{FRUCT}})}, Apr. 2022.

\bibitem{sobol1967distribution}
I.~M. Sobol', ``On the {{Distribution}} of {{Points}} in a {{Cube}} and the
  {{Approximate Evaluation}} of {{Integrals}},'' \emph{Zhurnal Vychislitel'noi
  Matematiki i Matematicheskoi Fiziki}, no.~4, 1967.

\bibitem{srinivas2010gaussian}
N.~Srinivas, A.~Krause, S.~Kakade, and M.~Seeger, ``Gaussian {{Process
  Optimization}} in the {{Bandit Setting}}: {{No Regret}} and {{Experimental
  Design}},'' in \emph{International {{Conference}} on {{Machine Learning}}
  ({{ICML}})}, Jun. 2010.

\bibitem{talbi2021automated}
E.-G. Talbi, ``Automated {{Design}} of {{Deep Neural Networks}}: {{A Survey}}
  and {{Unified Taxonomy}},'' \emph{Computing Surveys}, no.~2, Mar. 2021.

\bibitem{vasiljevic2021compute}
J.~Vasiljevic, L.~Bajic, D.~Capalija, S.~Sokorac, D.~Ignjatovic, L.~Bajic,
  M.~Trajkovic, I.~Hamer, I.~Matosevic, A.~Cejkov, U.~Aydonat, T.~Zhou, S.~Z.
  Gilani, A.~Paiva, J.~Chu, D.~Maksimovic, S.~A. Chin, Z.~Moudallal,
  A.~Rakhmati, S.~Nijjar, A.~Bhullar, B.~Drazic, C.~Lee, J.~Sun, K.-M. Kwong,
  J.~Connolly, M.~Dooley, H.~Farooq, J.~Y.~T. Chen, M.~Walker, K.~Dabiri,
  K.~Mabee, R.~S. Lal, N.~Rajatheva, R.~Retnamma, S.~Karodi, D.~Rosen,
  E.~Munoz, A.~Lewycky, A.~Knezevic, R.~Kim, A.~Rui, A.~Drouillard, and
  D.~Thompson, ``Compute {{Substrate}} for {{Software}} 2.0,'' \emph{IEEE
  Micro}, no.~2, Mar. 2021.

\bibitem{venkatesan2019magnet}
R.~Venkatesan, Y.~S. Shao, M.~Wang, J.~Clemons, S.~Dai, M.~Fojtik, B.~Keller,
  A.~Klinefelter, N.~Pinckney, P.~Raina, Y.~Zhang, B.~Zimmer, W.~J. Dally,
  J.~Emer, S.~W. Keckler, and B.~Khailany, ``{{MAGNet}}: {{A Modular
  Accelerator Generator}} for {{Neural Networks}},'' in \emph{International
  {{Conference}} on {{Computer-Aided Design}} ({{ICCAD}})}, Nov. 2019.

\bibitem{wilson2016deep}
A.~G. Wilson, Z.~Hu, R.~Salakhutdinov, and E.~P. Xing, ``Deep {{Kernel
  Learning}},'' in \emph{Artificial {{Intelligence}} and {{Statistics}}
  ({{AISTATS}})}, May 2016.

\bibitem{wistuba2020fewshot}
M.~Wistuba and J.~Grabocka, ``Few-{{Shot Bayesian Optimization}} with {{Deep
  Kernel Surrogates}},'' in \emph{International {{Conference}} on {{Learning
  Representations}} ({{ICLR}})}, Oct. 2020.

\bibitem{xi2020smaug}
S.~L. Xi, Y.~Yao, K.~Bhardwaj, P.~Whatmough, G.-Y. Wei, and D.~Brooks,
  ``{{SMAUG}}: {{End-to-End Full-Stack Simulation Infrastructure}} for {{Deep
  Learning Workloads}},'' \emph{Transactions on Architecture and Code
  Optimization (TACO)}, no.~4, Nov. 2020.

\bibitem{xiao2021hasco}
Q.~Xiao, S.~Zheng, B.~Wu, P.~Xu, X.~Qian, and Y.~Liang, ``{{HASCO}}: {{Towards
  Agile HArdware}} and {{Software CO-design}} for {{Tensor Computation}},'' in
  \emph{International {{Symposium}} on {{Computer Architecture}} ({{ISCA}})},
  Jun. 2021.

\bibitem{yang2020interstellar}
X.~Yang, M.~Gao, Q.~Liu, J.~Setter, J.~Pu, A.~Nayak, S.~Bell, K.~Cao, H.~Ha,
  P.~Raina, C.~Kozyrakis, and M.~Horowitz, ``Interstellar: {{Using Halide}}'s
  {{Scheduling Language}} to {{Analyze DNN Accelerators}},'' in
  \emph{Architectural {{Support}} for {{Programming Languages}} and {{Operating
  Systems}} ({{ASPLOS}})}, Mar. 2020.

\bibitem{yazdanbakhsh2020apollo}
A.~Yazdanbakhsh, C.~Angermueller, B.~Akin, Y.~Zhou, A.~Jones, M.~Hashemi,
  K.~Swersky, S.~Chatterjee, R.~Narayanaswami, and J.~Laudon, ``Apollo:
  {{Transferable Architecture Exploration}},'' \emph{Workshop on ML for
  Systems}, 2020.

\bibitem{zhang2022fullstack}
D.~Zhang, S.~Huda, E.~Songhori, K.~Prabhu, Q.~Le, A.~Goldie, and A.~Mirhoseini,
  ``A {{Full-Stack Search Technique}} for {{Domain Optimized Deep Learning
  Accelerators}},'' in \emph{Architectural {{Support}} for {{Programming
  Languages}} and {{Operating Systems}} ({{ASPLOS}})}, Feb. 2022.

\bibitem{zhuang2021comprehensive}
F.~Zhuang, Z.~Qi, K.~Duan, D.~Xi, Y.~Zhu, H.~Zhu, H.~Xiong, and Q.~He, ``A
  {{Comprehensive Survey}} on {{Transfer Learning}},'' \emph{IEEE}, no.~1, Jan.
  2021.

\end{thebibliography}

\end{document}